\documentclass{iau}
\usepackage{graphicx}
\usepackage{natbib}

\title[Solar magneto-convection] 
{Solar magneto-convection}

\author[Manfred Sch{\"u}ssler]   
{Manfred Sch{\"u}ssler}

\affiliation{Max-Planck-Institut f{\"ur} Sonnensystemforschung\\ 
37191 Katlenburg-Lindau, Germany\\ 
} 

\pubyear{2013}
\volume{294}  
\pagerange{xxx--xxx}
\date{\today}
\jname{Solar and Astrophysical Dynamos and Magnetic Activity}
\editors{A.G. Kosovichev, E.M. de Gouveia Dal Pino, Y.Yan,  eds.}
\begin{document}

\maketitle

\begin{abstract}
  An overview is given about recent developments and results of
  comprehensive simulations of magneto-convective processes in the
  near-surface layers and photosphere of the Sun. Simulations now cover
  a wide range of phenomena, from whole active regions, over
  individual sunspots and pores, magnetic flux concentrations and
  vortices in intergranular lanes, down to the intricate mixed-polarity
  structure of the magnetic field generated by small-scale dynamo action.
  The simulations in concert with high-resolution observations have
  provided breakthroughs in our understanding of the structure and
  dynamics of the magnetic fields in the solar photosphere.
  \keywords{Sun: magnetic fields, Sun: photosphere, convection,
  (magnetohydrodynamics:) MHD}
\end{abstract}

\firstsection 
\section{Introduction}
\label{sec:intro}

Solar magneto-convection is a rather extended area of research. It
covers topics ranging from the generation of the large-scale magnetic
field by the solar dynamo in the deep convection zone to the excitation,
propagation, and dissipation of currents and MHD waves in the
chromosphere and beyond. Obviously, this wide field cannot be reasonably
covered in a single review. Recent reviews of various aspects of solar
magneto-convection can be found in \citet{Fan:2009},
\citet{Charbonneau:2010}, \citet{Stein:2012}, and \citet{Weiss:2012}.

Here I shall restrict the discussion to the solar photosphere, the
region where the best understanding of the magneto-convective processes
has been achieved so far. The significant progress was made possible
through the interplay between high-resolution observations, idealized
theoretical models, and comprehensive numerical simulations. The latter
provide a 3D view of the processes underlying the
observations. Simulations can be validated (albeit not in a strict
sense) by detailed comparison of synthetic observational quantities
based on simulation results with real observations. At the same time,
this comparison provides a firm basis for the interpretation of the
observations in terms physical quantities and processes.

I prefer to qualify these simulations as `comprehensive' in favor of the
often-used term `realistic', the reason being that the values of
important non-dimensional numbers (such as the kinetic and magnetic
Reynolds numbers, or the magnetic Prandtl number) that can be reached in
the simulations differ from the realistic values by (many) orders of
magnitude. Typically, some kind of hyperdiffusion or subgrid modelling
is used to account for the unresolved spatial scales, but in the absence
of a theory of turbulence the validity of such quasi-heuristic
approaches cannot be strictly proven. On the other hand, the term
`comprehensive' expresses the fact that these simulations aim at
representing all relevant physical processes and conditions in the
photosphere, i.e., compressible MHD in 3D domain, partial ionization and
molecule formation, as well as non-grey and non-local radiative
transfer. A number of numerical codes are used to obtain comprehensive
simulations, among them the ANTARES code \citep{Muthsam:etal:2010}, the
CO5BOLD code \citep{Freytag:etal:2011}, the MURaM code
\citep{Voegler:2003, Voegler:etal:2005}, the PENCIL code
\citep{Brandenburg:Dobler:2002}, the STAGGER/BIFROST code
\citep{Gudiksen:etal:2011}, as well as the codes of
\citet{Robinson:etal:2003}, and \citet{Jacoutot:etal:2008}.  There are
also codes with rather coarse treatment of radiation, such as the RADMHD
code \citep{Abbett:2007, Abbett:Fisher:2012} with a local radiative
cooling term or the code of \citet{Ustyugov:2010}, which uses the
diffusion approximation.  Unfortunately, not all groups have given
detailed descriptions of the numerical methods and approximations used
in their codes and shown test results in the literature, so that it is
difficult to judge and trust the validity of their simulation
results. On the other hand, the STAGGER, CO5BOLD and MURaM groups groups
have embarked on a cross-validation of their codes by a detailed
comparison of numerical results \citep{Beeck:etal:2012}.

Even when restricting oneself to the photosphere, a complete review of
observational, theoretical, and simulation results is far beyond the
scope of this contribution. Limitations of space (and human time) force
me to restrict myself mainly to the discussion of recent simulation
results, with very limited references to observation and theoretical
concepts. Furthermore, I am afraid that this text is somewhat biased
towards results obtained with the MURaM code -- not because this code
would be better than others, but just because I am most familiar with
its results. What follows is organized according to the spatial scale of
the phenomenon: from whole active regions down to the fine threads of
the magnetic field generated by small-scale dynamo action.

\section{Active regions}
\label{sec:ar}

In the past decade, the computational resources available for
comprehensive simulations have increased dramatically. It has become
possible to simulate the emergence and development of whole active
regions with sunspots, pores, and plage areas. This line of research was
started by \citet{Cheung:etal:2007b}, who considered the emergence of a
twisted flux tube carrying a longitudinal magnetic flux of
$10^{19}\,$Mx. The initially coherent flux tube becomes undulated and
fragmented by the vigorous convective flows, so that the flux emergence
takes place in the form of numerous patches of horizontal field
connecting small bipoles in a `salt-and-pepper' pattern. The disturbed
granulation at the emergence site shows elongated granules
threaded by dark lanes. Many of these features are also found in
observations \citep[e.g.,][]{Cheung:etal:2008, Yelles:etal:2009}.
Similar results were found by \citet{Stein:etal:2011},
\citet{Fang:etal:2010, Fang:etal:2012}, and by
\citet{Martinez:etal:2008, Martinez:etal:2009}. The latter authors also
considered the evolution of the coronal field during flux emergence and
the development of spicule-like structures.

\citet{Cheung:etal:2010} simulated the emergence of a flux tube with
$7.6\times 10^{21}\,$Mx.  As the emergence proceeds, the pattern of
small bipoles at the visible solar surface increasingly reorganizes
until the magnetic flux distribution reflects the large-scale bipolar
structure of the underlying Omega-shaped flux tube. At the same time, a
pair of sunspots forms through the coalescence of pores and smaller flux
fragments, so that the appearance of a full-fledged active region
develops. These simulations also solved the long-standing problem of how
the enormous amount of mass contained in a deep-lying flux tube is
removed as it reaches the low-density regions near the surface. It turns
out that much of this mass is accumulated in sinking U-shaped loops that
reconnect near the surface and thus release the mass from the field
lines of the surface field. Observational evidence for this process has
recently been reported \citep{Centeno:2012}.  Simulations by M. Cheung
and M. Rempel (not yet published) show that an initial twist is not
essential for the formation of coherent sunspots from a flux tube
inserted at a depth of about 16~Mm.

While the simulations discussed so far assumed a large-scale flux tube
to coherently enter the simulation domain from below,
\citet{Stein:Nordlund:2012} assumed horizontal magnetic field to be
advected by upflows at the bottom boundary of the computational
domain. They found the development of a bipolar structure at the surface
that resembled a small active region with a few pores. Although this
simulation shows that a bipolar region can form without inserting a
coherent flux tube beforehand, some caution is indicated when
interpreting this result: the horizontal field advected into the
computational box has a fixed direction and strength, so that the
simulation implicitely assumes the existence of an infinite reservoir of
strictly organized and coherent horizontal magnetic flux below the
computational domain, in fact something like a huge homogeneous flux
sheet. Consequently, the orientation of the developing bipolar region is
along the direction of the assumed horizontal field below the simulation
box.

\section{Sunspots and pores}
\label{sec:spots}

Comprehensive simulations of sunspots have seen tremendous progress in
the last few years. Already studies of `sunspot slabs' at rather coarse
resolution and limited box size \citep{Heinemann:etal:2007,
Rempel:etal:2009b} provided crucial insights in the processes
responsible for the penumbral filamentation and the Evershed effect. The
first simulations of full sunspots by \citet{Rempel:etal:2009a} and
their further development and detailed analysis by \citet{Rempel:2011a,
Rempel:2011b, Rempel:2012a} revealed the nature of the
magneto-convective processes leading to the characteristic structure of
sunspots. While the bright umbral dots represent hot upflow plumes,
whose lateral expansion is strongly constricted by the vertical umbral
field \citep{Schuessler:Voegler:2006, Bharti:etal:2010}, the inclined
field in the penumbra provides a path for the unimpeded expansion of hot
upflowing material: the direction along the field. As a consequence,
sheet-like convective upflows in the penumbra appear as bright,
elongated filaments with strong outflows (representing the Evershed
effect), which feed downflow patches in the outer part of the
penumbra. Part of the rising plasma turns over laterally (perpendicular
to the filament axis) and descends in downflow lanes next to the bright
filaments. The expanding outflow stretches and expels the magnetic
field, so that the outflow channels exhibit relatively weak (albeit
non-vanishing) and almost horizontal magnetic field. All these
properties are consistent with observational results
\citep[e.g.,][]{Borrero:Ichimoto:2011, Rempel:Schlichenmaier:2011,
Joshi:etal:2011, Scharmer:etal:2011}.

It is observationally well established that the formation of big
sunspots invariably is connected to major events of magnetic flux
emergence. This is also reflected in simulations of sunspot formation
and structure, which always require either the advection of
well-organized magnetic flux into the computational domain (as in the
simulations of active-region formation discussed in Sec.~\ref{sec:ar})
or the assumption of a sufficiently big concentration of vertical
magnetic flux as initial condition for the simulation.  The situation is
different with pores, small dark flux concentrations without penumbrae
and mostly not exceeding the area of a few granules. Such structures are
observed to form and decay quasi-spontaneously in mature plage areas,
without any associated emergence of new magnetic flux.
Magneto-convection simulations in sufficiently deep boxes also show the
formation of pore-like dark structures if enough magnetic flux is
available, either in the form of a preset vertical flux through the
computational box \citep{Kitiashvili:etal:2010, Ustyugov:2010} or by the
perpetual advection of horizontal magnetic flux through the bottom
boundary \citep{Stein:etal:2011}. Previous simulations in shallow boxes
extending to a few Mm depth below the visible solar surface did not show
the formation of dark structures, apart from occasional small
`micropores' \citep[e.g.,][]{Bercik:etal:2003, Voegler:etal:2005,
Stein:Nordlund:2006}. The reason why shallow simulations miss the
formation of proper pores is that it requires horizontal flows on a
sufficiently big spatial scale in order to collect the necessary
sufficient amount of magnetic flux.  Since the horizontal scale of the
convective flows increases with depth \citep[see, e.g., Fig.~6
in][]{Nordlund:etal:2009}, only sufficently deep boxes (at least 6 Mm)
provide the conditions for pore formation.

\begin{figure}[t!]
\begin{center}
 \includegraphics[width=6.5cm]{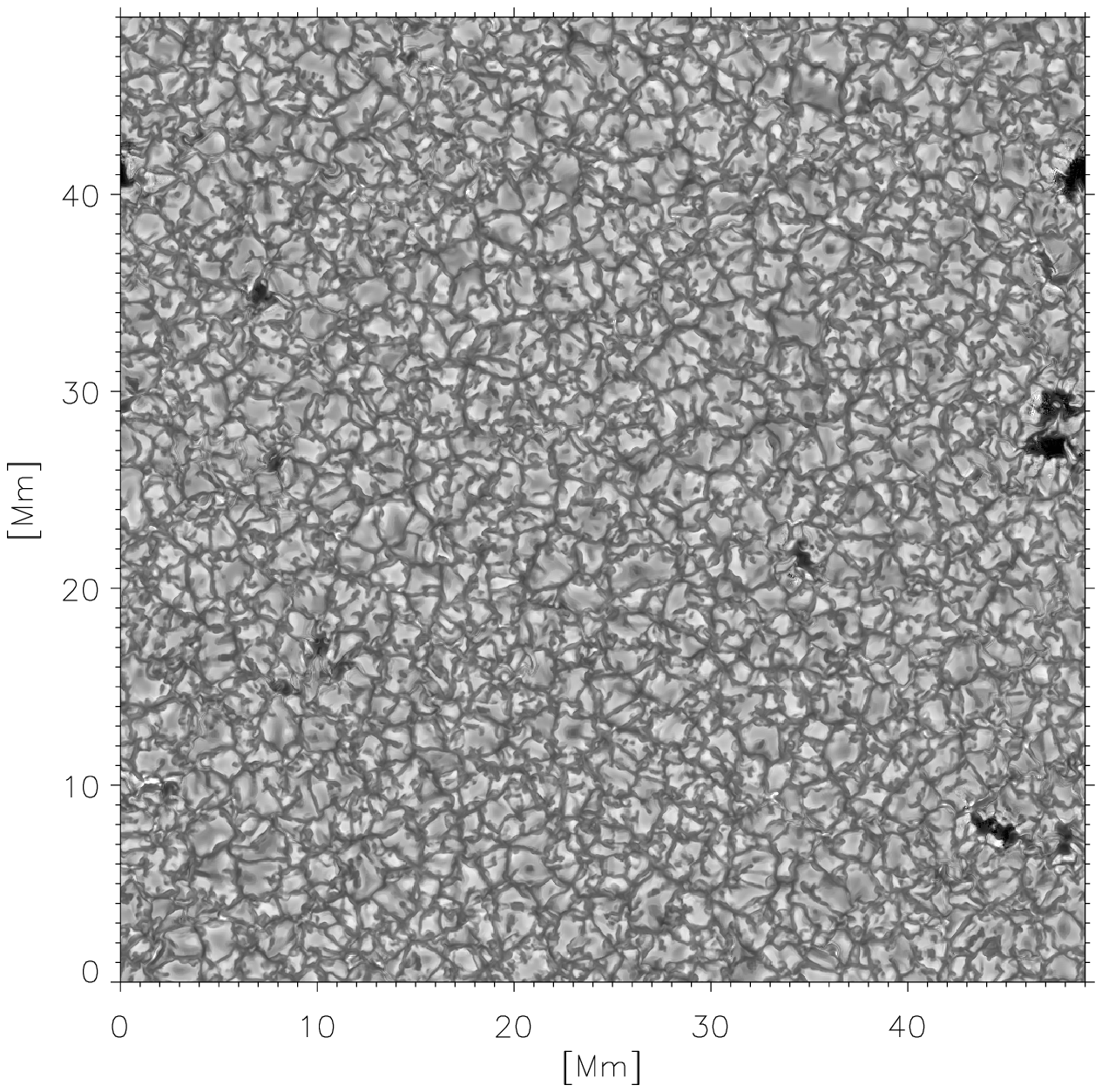}
 \includegraphics[width=6.5cm]{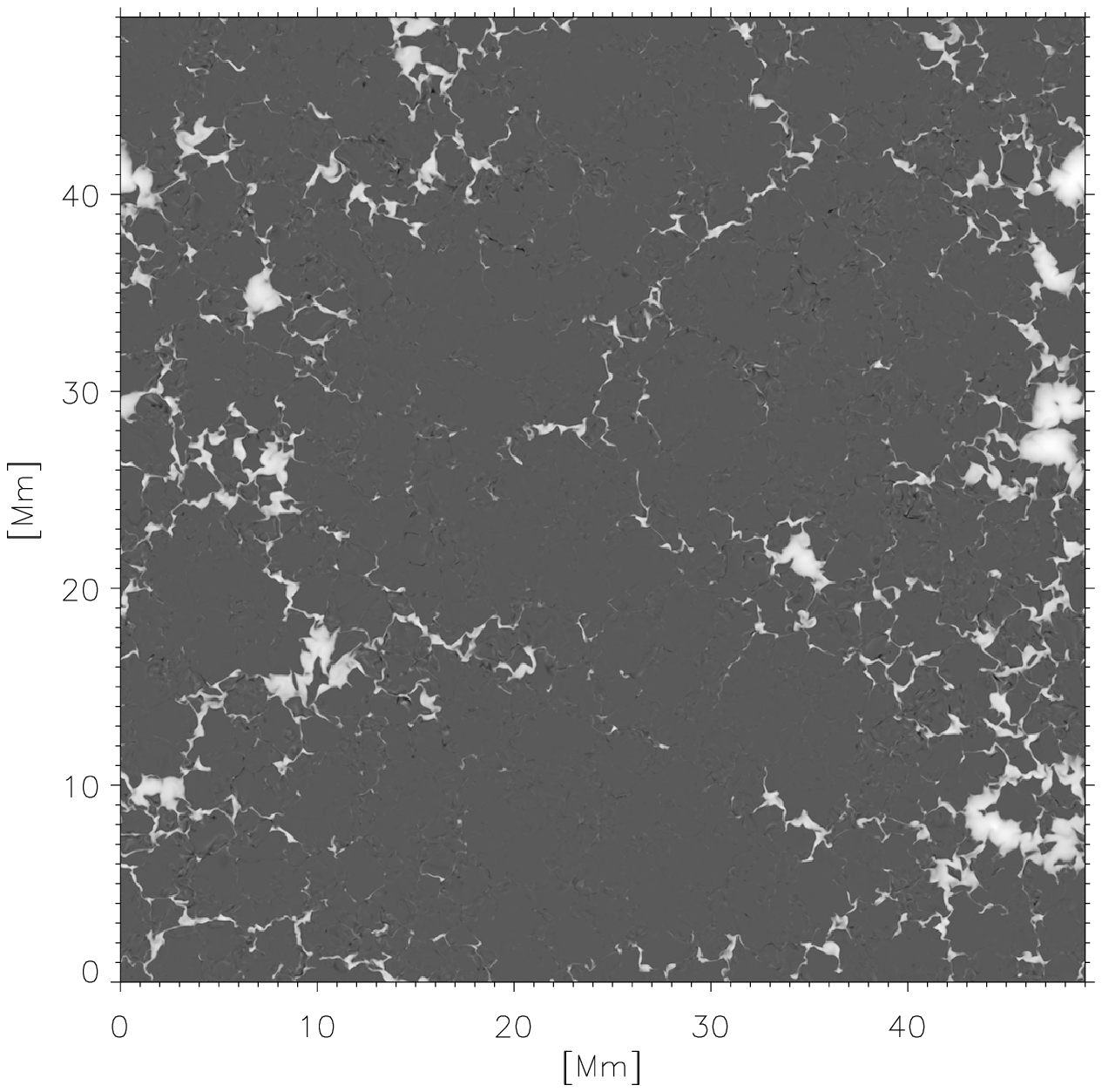}
 \caption{Maps of brightness (bolometric intensity, left panel) and
 vertical magnetic field at the optical solar surface (right panel) from
 a snapshot of a magneto-convection simulation in a $49\times49\,$Mm$^2$
 wide and $15\,$Mm deep box permeated by a horizontally averaged
 vertical field of 100~G.}
   \label{fig:plage}
\end{center}
\end{figure}

A word of caution is in order here. The maximum horizontal scale of the
convective motions seen in simulations is roughly equal to the
depth. This means that the horizontal extension of the computational box
should be at least 3-4 times its depth in order to provide enough volume
for a development of a large-scale flow pattern that is not restrained
by the periodic bounday conditions at the side walls of the
box. Simulation in boxes with smaller aspect ratios most probably
provide unrealistic results.

What determines the size (or amount of flux) of pores and why do big
sunspots not form in this way? This depends basically on the amount of
magnetic flux available (mean flux density) and on the relation of the
advection time of the horizontal flow (horizontal scale divided by flow
speed) and the lifetime of the flux concentration: the advected flux per
unit time has to be at least equal to the flux lost by `turbulent
erosion' of the flux concentration \citep{Cameron:etal:2007b}. As a
result, for the mean vertical flux densities found in plages (a few
hundred G), the biggest flux concentrations formed by flux advection do
not exceed the size of a few granules. As an example,
Fig.~\ref{fig:plage} shows maps of the emergent intensity (left panel)
and of the surface distribution of vertical magnetic field (right panel)
from a simulation (run by M.C.M. Cheung) in a box of
$49\times49\,$Mm$^2$ horizontal extension reaching down to $15\,$Mm
depth, so that nearly supergranular-size flows can develop. The average
vertical field strength is 100~G, thus representing a typical plage
region.  The simulation formed a number of pores, which
qualitatively resemble those occuring in plage areas. Furthermore,
the distribution of the magnetic field shows a multi-cellular
pattern, reflecting the various horizontal scales of the convective
flow. The biggest cells have a scale of about 15~Mm, corresponding to
the depth of the computational box. This pattern is remarkably similar
to those seen in plage observations.  This result opens the 
possibility to obtain information about the flows in the upper part of
the convection zone by surface observations and comparison with
simulations, providing a supplement to helioseismic
inversions.

\begin{figure}[ht!]
\begin{center}
 \includegraphics[width=\linewidth]{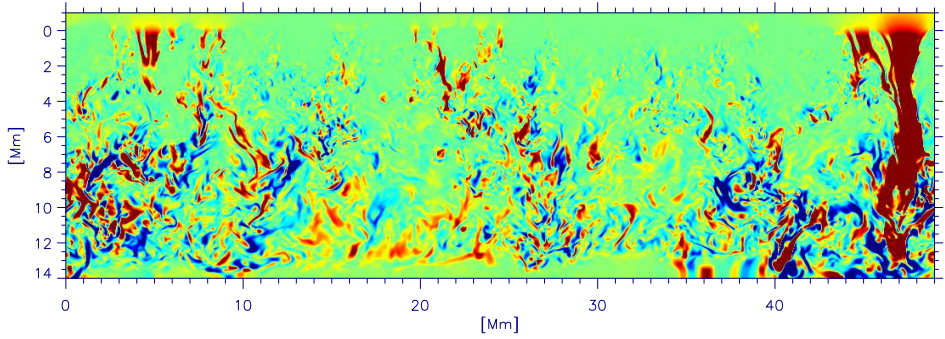}
 \caption{Map of the vertical magnetic field component, $B_z$, in a
 vertical cut along a horizontal line at $x=27.5\,$Mm in
 Fig.~\ref{fig:plage} (right panel).  Reddish colors indicate positive
 values of $B_z$, bluish colors negative values. The zero level of the
 depth scale corresponds to the optical surface (mean level of optical
 depth unity).}
   \label{fig:plage_cut}
\end{center}
\end{figure}

Fig.~\ref{fig:plage_cut} shows the distribution of the vertical magnetic
field component on a vertical section of the simulation box. The section
cuts through surface flux concentrations of various sizes. While the
small-scale flux concentrations in intergranular lanes are restricted to
the near-surface layers, bigger flux concentrations extend deeper,
reflecting the depth of the flow pattern that accumulated them in the
first place. The big magnetic flux concentration corresponding to the
pore near the right-hand edge of the plot stretches down almost to the
bottom of the computational domain. While the magnetic field is rather
organized in the surface layers, the deeper parts of the box show
patches with a chaotic pattern of small-scale field structures,
resulting from the turbulent nature of the convective downflow regions.
These are the sites of small-scale dynamo action (see
Sec.~\ref{sec:dynamo}).

\section{Small-scale flux concentrations}
\label{sec:flux_conc}

Starting with the pioneering work of \citet{Nordlund:1983,
Nordlund:1985b}, comprehensive 3D simulations have revealed the
processes responsible for the concentration of most of the
`non-turbulent' part of the magnetic flux at the solar surface into
small-scale structures with kG field strength located in intergranular
downflow lanes \citep[e.g.,][]{Steiner:etal:1996, Steiner:etal:1998,
Bercik:etal:1998, Stein:etal:2002, Voegler:Schuessler:2003,
Voegler:etal:2005, Schaffenberger:etal:2006, Stein:Nordlund:2006}. The
simulations confirmed the validity of the basic theoretical explanations
for the intermittent structure of the solar surface field, which were
developed already in the 60s and 70s of the last century. The first
concept is {\em flux expulsion} by convective flows in an electrically
conducting fluid, pioneered by \citet{Weiss:1964, Weiss:1966} and
\citet{Parker:1963}.  The advection of magnetic flux by horizontal flows
and its concentration in convective downflow areas concentrates the
field roughly up to equipartition field strength, for which the magnetic
energy density equals the kinetic energy density of the flow. The
concept of {\em convective collapse} is based in the strongly
superadiabatic startification of the subsurface layers, which renders
equipartition flux concentrations unstable with respect to downflow
along the field lines \citep{Webb:Roberts:1978,
Spruit:Zweibel:1979}. This evacuates the upper part of the flux
concentrations, so that the gas pressure of the surrounding plasma
compresses the flux concentration until kG field strength is reached
\citep{Parker:1978, Spruit:1979}.  The depression of the surface of
optical depth unity in the evacuated flux concentration leads to lateral
heating of its interior by the hot walls and its appearance as a bright
structure \citep{Spruit:1976, Deinzer:etal:1984b}.  Quantitative
comparison of observations and comprehensive numerical simulations
confirmed the relation between kG magnetic flux concentrations and
bright points \citep{Schuessler:etal:2003, Shelyag:etal:2004,
Lagg:etal:2010, Roehrbein:etal:2011, Danilovic:etal:2012}, faculae
\citep{Keller:etal:2004, Carlsson:etal:2004}, and the
spectro-polarimetric signatures of these structures
\citep{Shelyag:etal:2007}.

\section{Vortices}
\label{sec:vortex}

\begin{figure}[ht!]
\begin{center}
 \includegraphics[width=\linewidth]{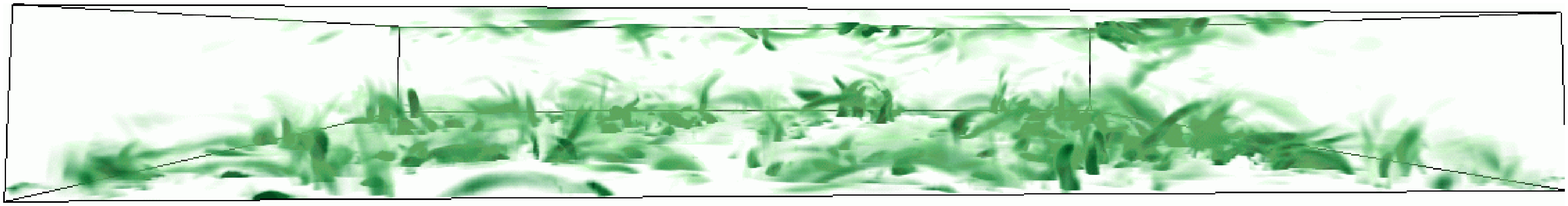}\\
 \includegraphics[width=\linewidth]{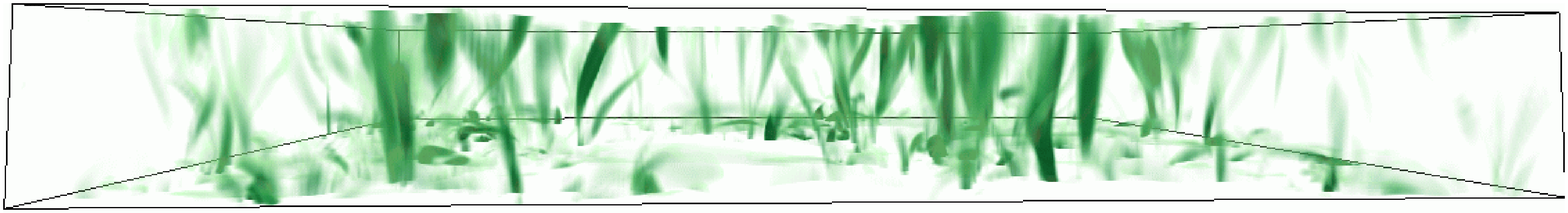}
 \caption{Volume rendering of the swirling strength \cite[for a
     definition, see][]{Moll:etal:2011b} for a non-magnetic simulation
     (upper panel) and for a simulation with an average vertical
     magnetic field of 200~G (lower panel), viewed from one side of the
     computational box. Shown is the upper half
     ($6\times6\times0.8\,{\rm Mm}^3$) of the box, the bottom plane
     corresponding roughly to the optical surface. While the vertical
     vortex tubes turn over near the surface in the non-magnetic case,
     they align with the vertical flux concentration and protrude high
     into the atmosphere when a background magnetic field is present.}
   \label{fig:vortex}
\end{center}
\end{figure}

The formal equivalence of the vorticity equation and the induction
equation suggests that vertical vorticity should be expelled and
concentrated in intergranular downflows in similar ways as magnetic
flux. It is no surprise, therefore, that vortical features and whirl
flows are regularly found in comprehensive convection simulations
\citep{Nordlund:1985a, Stein:Nordlund:1998, Moll:etal:2011b,
  Kitiashvili:etal:2012b}. While most of the strong vertical vortices
have diameters below 100~km, probably too small
to be detected with existing telecopes, larger-scale vortex flows
have been reported by observers \citep[e.g.,][]{Brandt:etal:1988,
  Bonet:etal:2008, Bonet:etal:2010, Attie:etal:2009,
  Vargas:etal:2011}. In addition, \citet{Steiner:etal:2010} detected
horizontal vortices in granules, both in simulations and in
observations with the {\em Sunrise} balloon telescope.

\begin{figure}[ht!]
\begin{center}
 \includegraphics[width=0.30\linewidth]{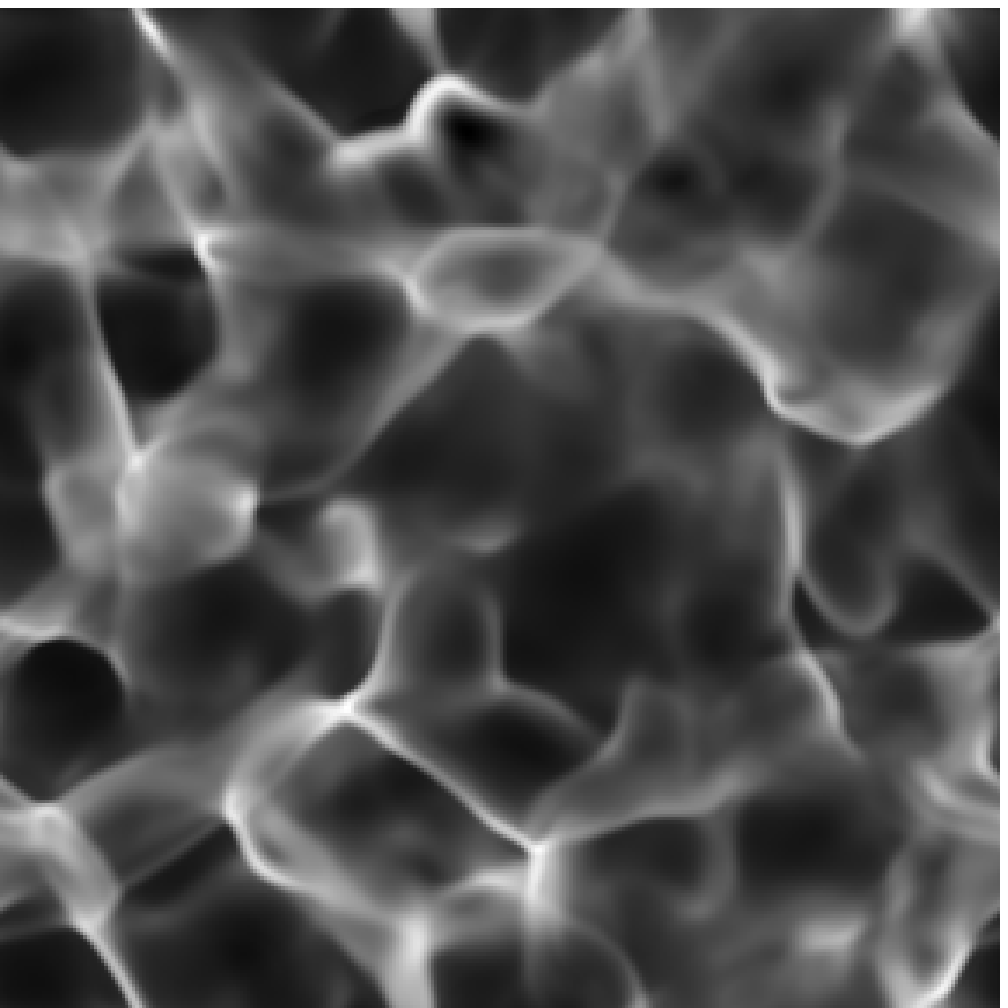}\hspace{2mm}
 \includegraphics[width=0.30\linewidth]{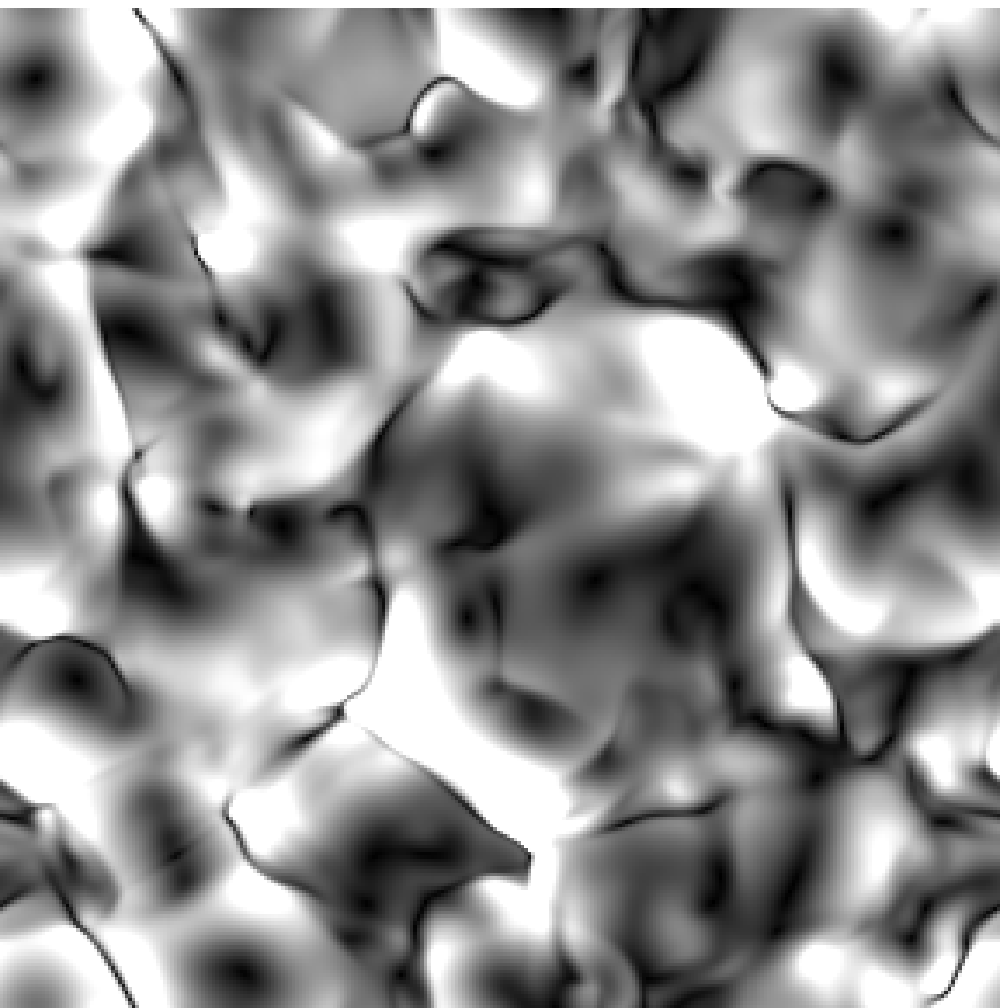}\hspace{2mm}
 \includegraphics[width=0.30\linewidth]{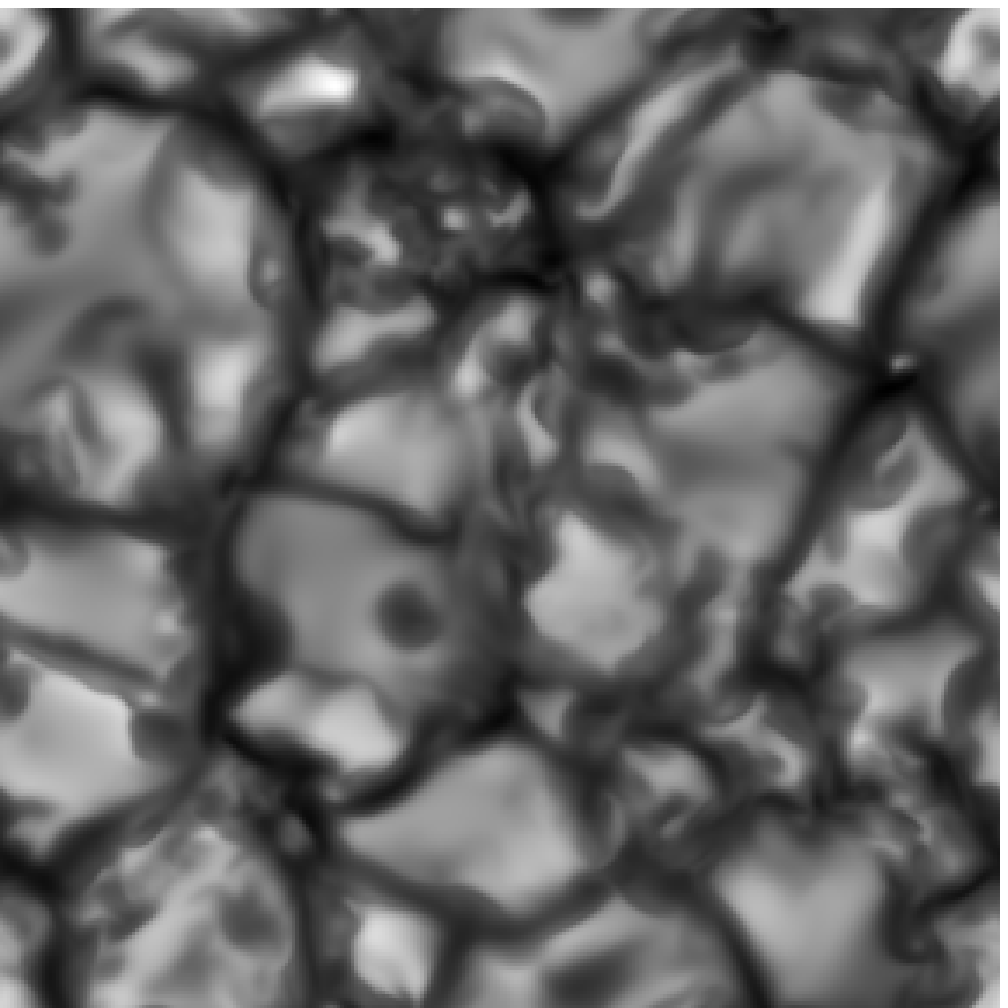}\vspace{3mm}\\
 \includegraphics[width=0.30\linewidth]{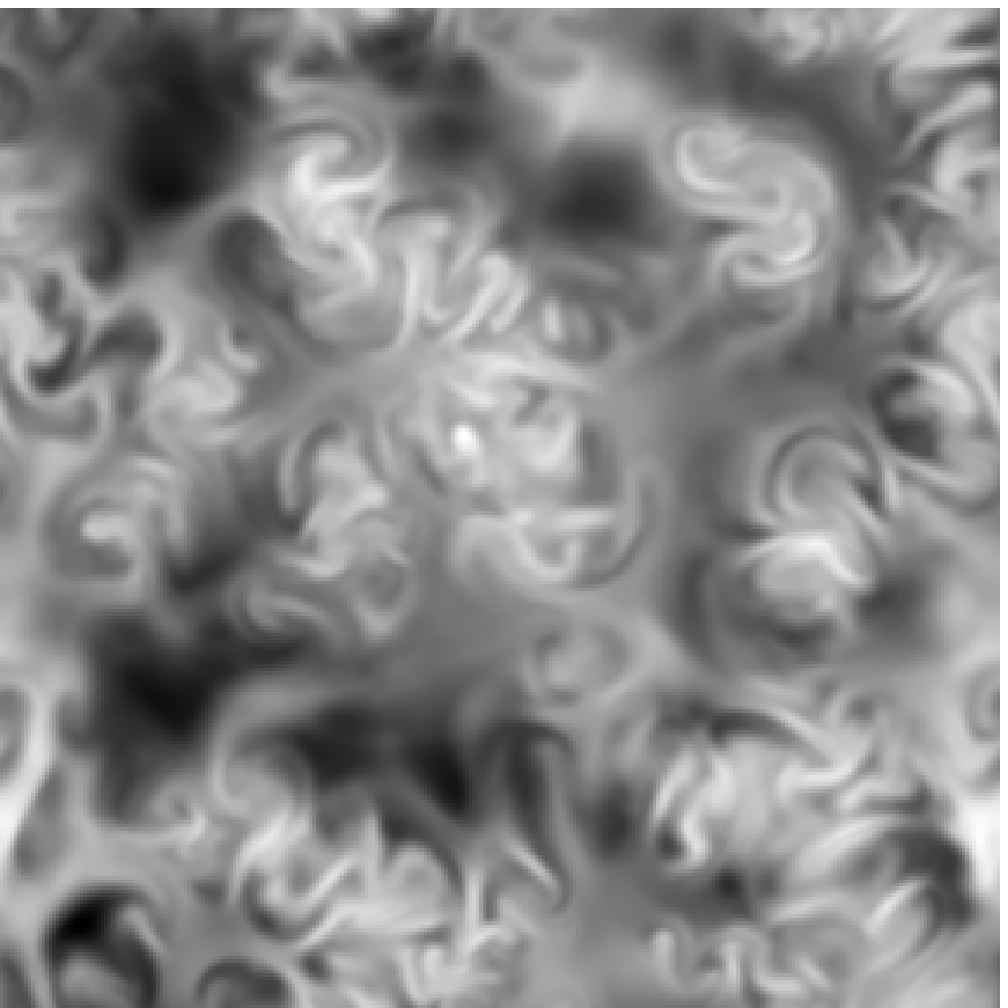}\hspace{2mm}
 \includegraphics[width=0.30\linewidth]{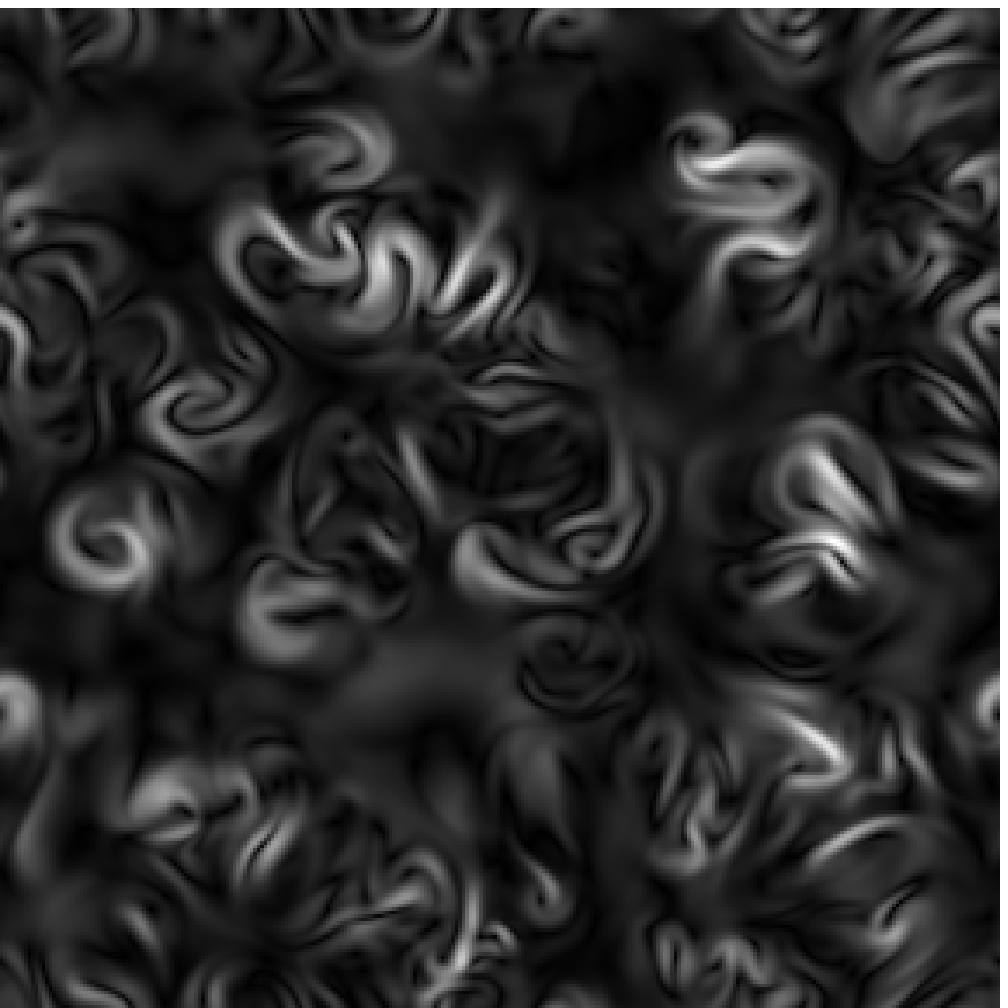}\hspace{2mm}
 \includegraphics[width=0.30\linewidth]{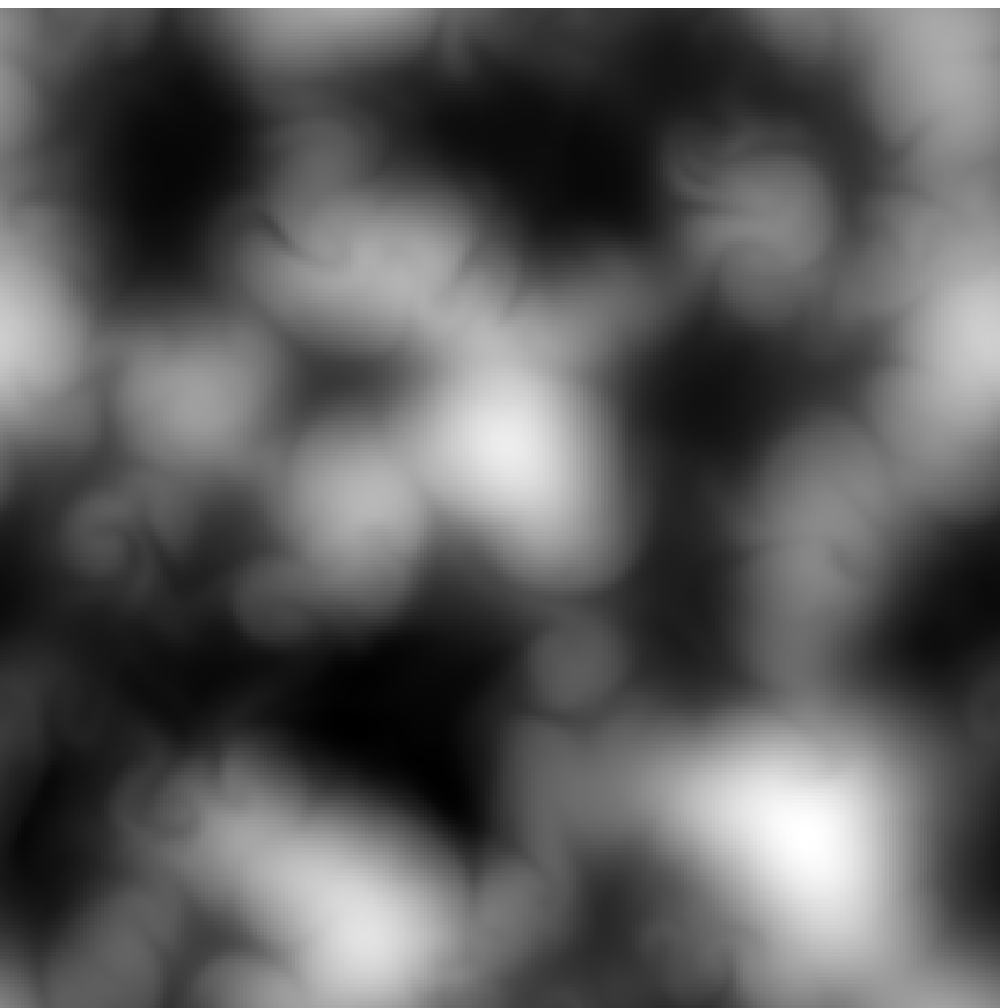}
 \caption{Horizontal cuts (size $6\,{\rm Mm}\times 6\,{\rm Mm}$) through
   the upper photospheric layers of a nonmagnetic simulation (upper row)
   and a simulation with an average vertical magnetic field of 200~G
   (lower row). Shown are maps of temperature (left panels), ranging
   from about 3000~K to over 7000~K, and of horizontal flow speed
   (middle panels), which reaches values up to 15~km$\cdot$s$^{-1}$).
   The upper right panel shows a (bolometric) brightness map for the
   hydrodynamic case and the lower right panel gives a map of the
   vertical magnetic field in the upper photosphere of the magnetic
   run. The hydrodynamic case is dominated by shock fronts while
   vortices prevail in the magnetic simulation. Both kinds of structures
   show strong local heating of the plasma.}
   \label{fig:cuts}
\end{center}
\end{figure}

If a sufficient amount of vertical background flux is present, the
vortices are closely associated with magnetic flux
concentrations. \citet{Voegler:2004} first found `magnetic vortices' in
magneto-convection simulations. Recently, this topic found considerable
interest, also in connection with chromospheric and coronal heating
processes \citep[e.g.,][]{Shelyag:etal:2011, Kitiashvili:etal:2012a,
Steiner:Rezaei:2012, Wedemeyer:etal:2012a}.  The simulations of
\citet{Moll:etal:2012} indicate that the dynamical structure of the
upper photosphere and chromosphere of `quiet' regions (i.e., areas with
a mean signed field below about {20}~G) is significantly different from
that of more strongly magnetized regions: while the former are dominated
by patterns of moving shock fronts \citep[as noted previously
by][]{Wedemeyer:etal:2004, Wedemeyer:2010}, the more magnetized areas
show numerous vertical vortices, which are strongly associated with the
small-scale magnetic flux concentrations in intergranular lanes.  As
illustrated in Fig.~\ref{fig:vortex}, vertically orientated vortices in
the surface layers of weakly magnetized regions do not extend high into
the atmosphere, but turn over near the optical surface and form
low-lying vortex loops \citep{Moll:etal:2011b}. In more strongly
magnetized regions (network, plage), the vertical vortices align with
the magnetic flux concentrations and reach high into the atmosphere,
where the dissipation of kinetic energy leads to intense local
heating. Fig.~\ref{fig:cuts} illustrates the different dynamical
character of the upper photosphere/lower chromosphere in weakly and
strongly magnetized regions: while shocks dominate the former, hot
vortices provide the governing structure in the latter case.

\section{Small-scale dynamo}
\label{sec:dynamo}

There are a number of observational indications for the existence of a
considerable amount of magnetic flux in the form of a small-scale
turbulent field in the solar near-surface layers \citep[][see also the
contribution of S.~Tsuneta in this volume]{Sanchez:Martinez:2011}. This
component of the solar surface flux seesm to be independent of the
11-year solar cycle \citep[e.g.,][]{Trujillo:etal:2004}, indicating a
different source mechanism. Recently, D.~B{\"u}hler (in preparation)
systematically analyzed Hinode/SP data for the period between 2007 and
2012, which covers the extended solar minimum and the rise of the current
cycle. He found that the area fraction of weak, but significant, linear
and circular polarization signals in quiet internetwork regions did not
change during this time. This suggests that the turbulent magnetic field
is not connected to the solar cycle, but possibly generated by
small-scale dynamo (SSD) action.

Characteristic for a SSD driven by a turbulent flow in an electrically
conducting fluid is that the magnetic field is generated at spatial
scales that are much smaller than that of the energy-carrying eddies of
the flow (i.e., the integral scale of the turbulence). Although there is
some dependence on the value of the magnetic Prandtl number (ratio of
kinematic viscosity to magnetic diffusivity), SSD action probably occurs
in all turbulent flows of sufficiently high magnetic Reynolds number
\citep[see][and references therein]{Brandenburg:etal:2012}.  While
direct numerical simulations demonstrated SSD action in various settings
since the 1980s \citep[e.g.,][]{Meneguzzi:etal:1981, Cattaneo:1999,
Bushby:etal:2012}, the effect was also found in large-eddy simulations
(Boussinesq or anelastic) in spherical shells carried out to model the
solar convection zone \citep{Gilman:Miller:1981, Glatzmaier:1985,
Brun:etal:2004}.

Comprehensive simulations of solar near-surface convection indicate that
the observed turbulent field could indeed be a product of a SSD action
driven by the turbulent intergranular downflows
\citep{Voegler:Schuessler:2007, Pietarila:etal:2010, Moll:etal:2011a}.
The characteristic properties of a magnetic field generated by a SSD can
explain the observed strong horizontal fields observed in the middle
photosphere \citep{Schuessler:Voegler:2008} and the weak signals
detected with sensitive polarimeters \citep{Pietarila:etal:2009,
Danilovic:etal:2010a, Danilovic:etal:2010b}.  While most of the field
due to the SSD is of mixed polarity on small scales and has a strength
of some tens to a few hundred Gauss, occasionally enough unipolar flux
is being assembled by the granular flows to form a kG flux concentration
appearing as a bright point in the visible light. Therefore,
observations of such features in quiet internetwork areas on the Sun are
consistent with SSD action.

\begin{figure}[ht!]
\begin{center}
 \includegraphics[width=0.30\linewidth]{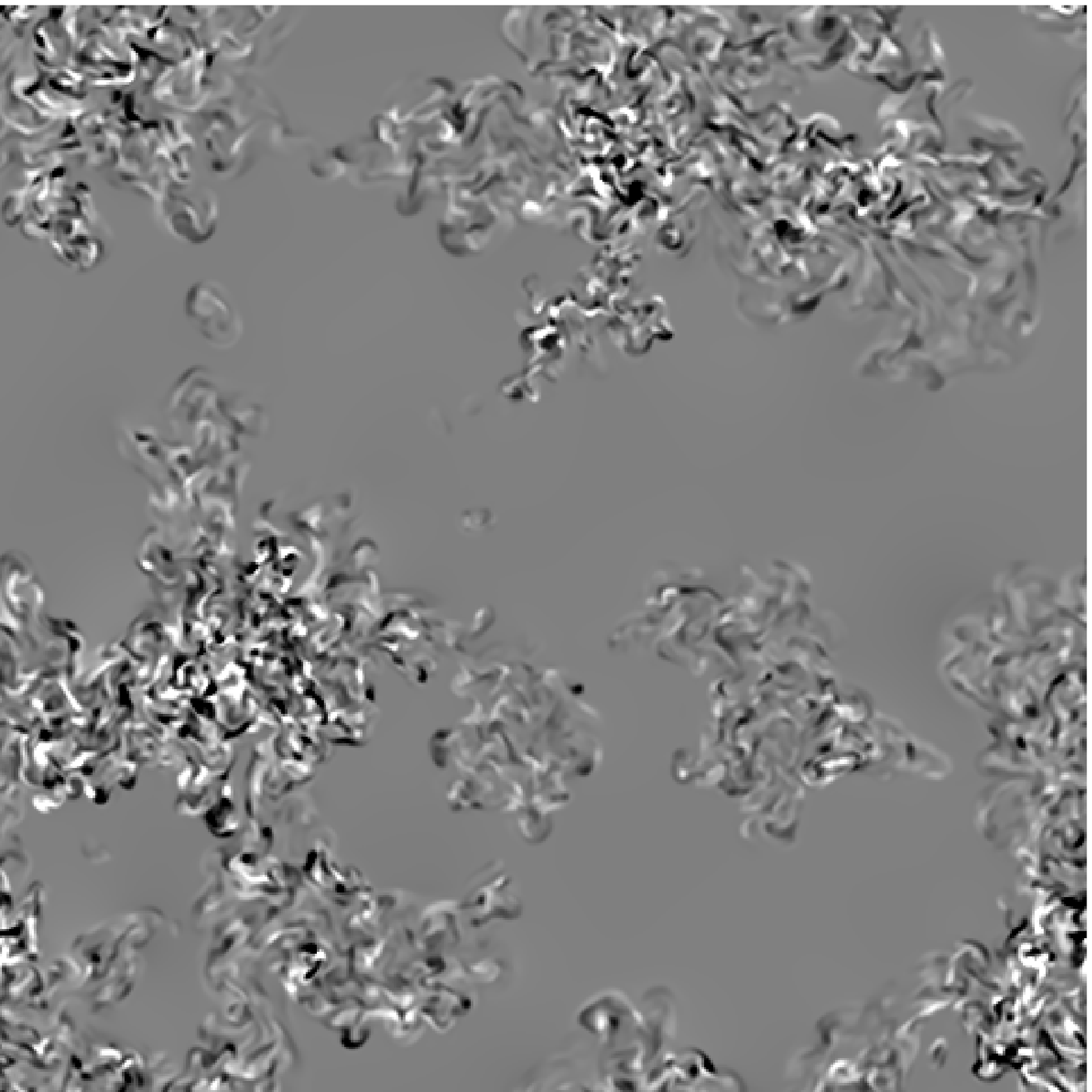}\hspace{2mm}
 \includegraphics[width=0.30\linewidth]{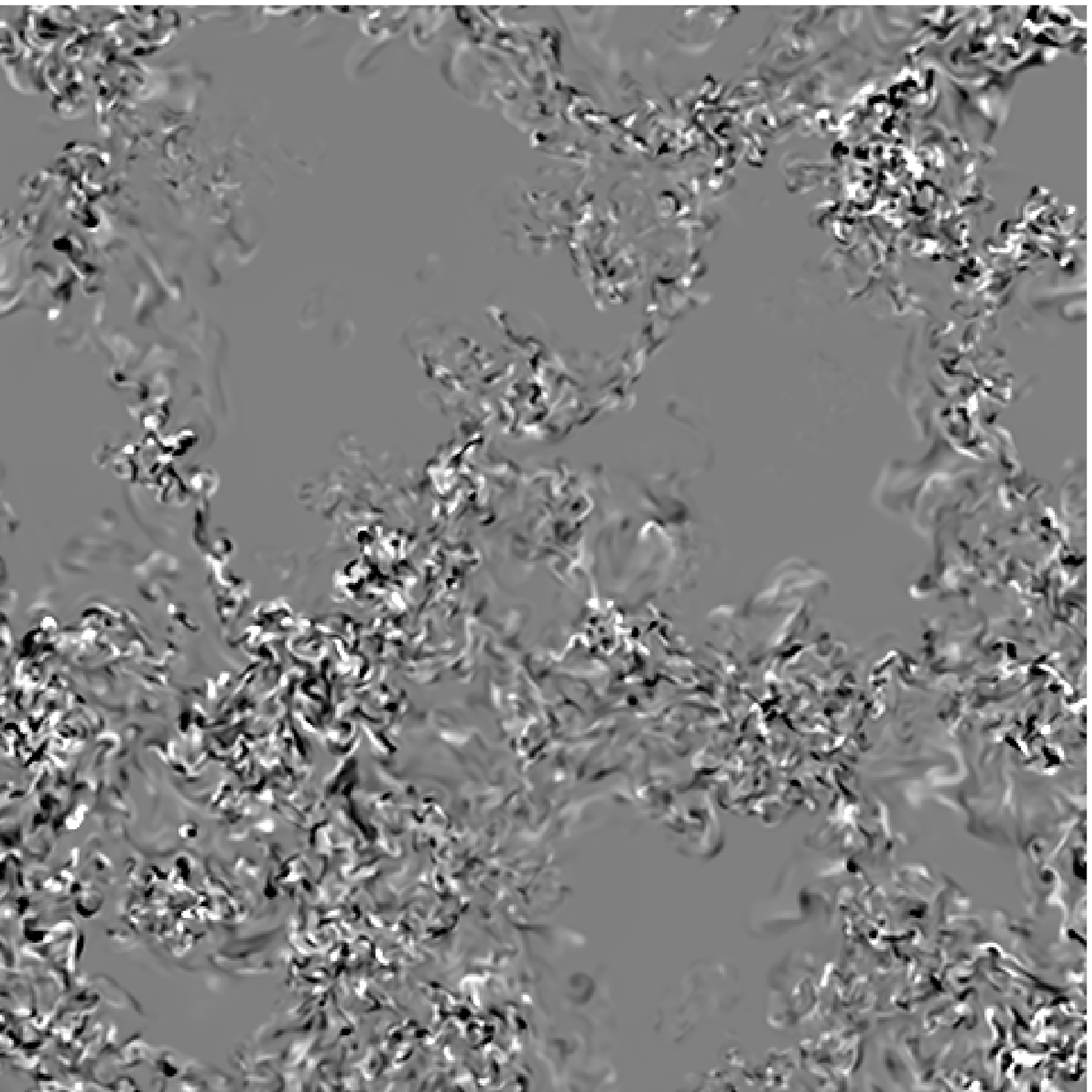}\hspace{2mm}
 \includegraphics[width=0.30\linewidth]{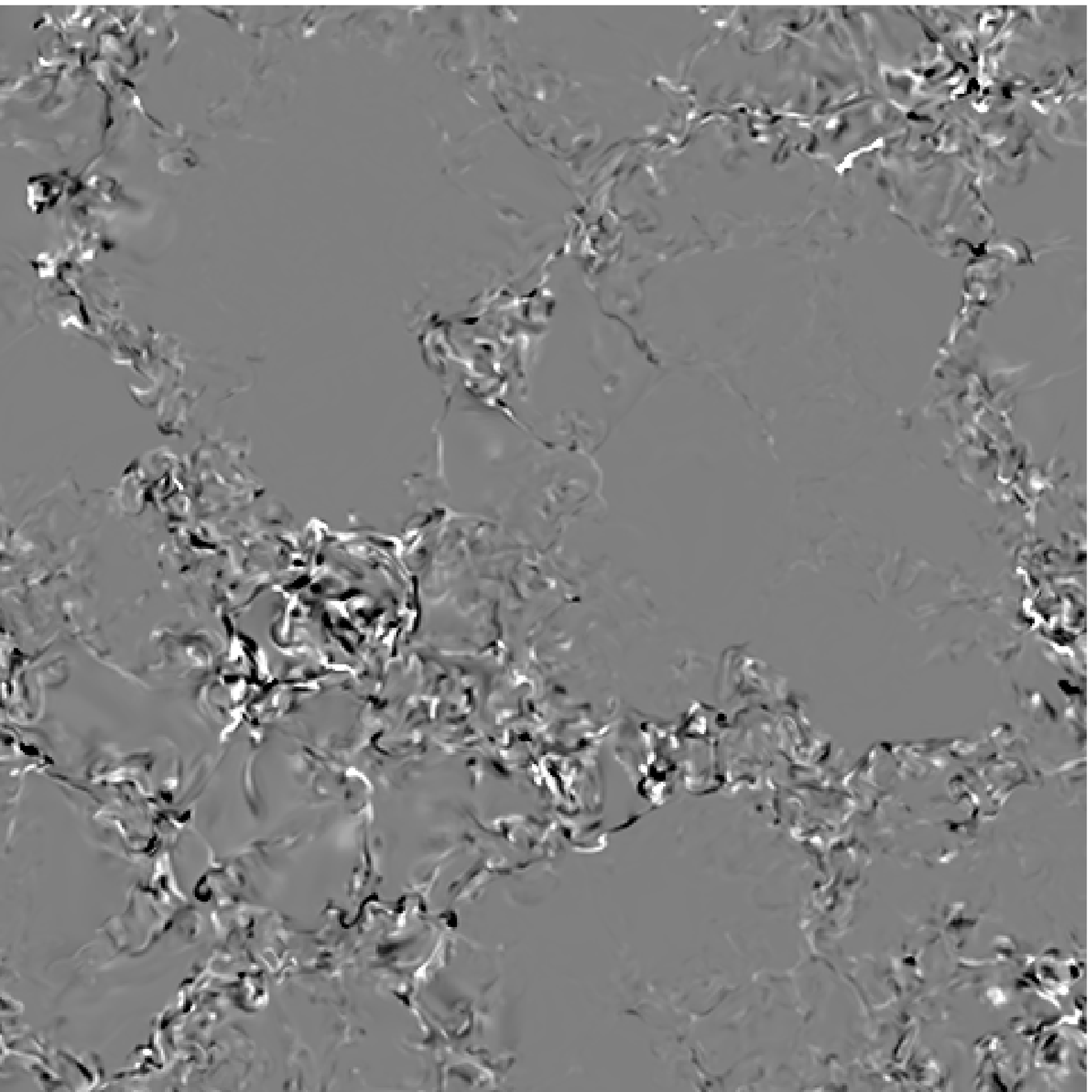}\vspace{3mm}\\
 \includegraphics[width=0.30\linewidth]{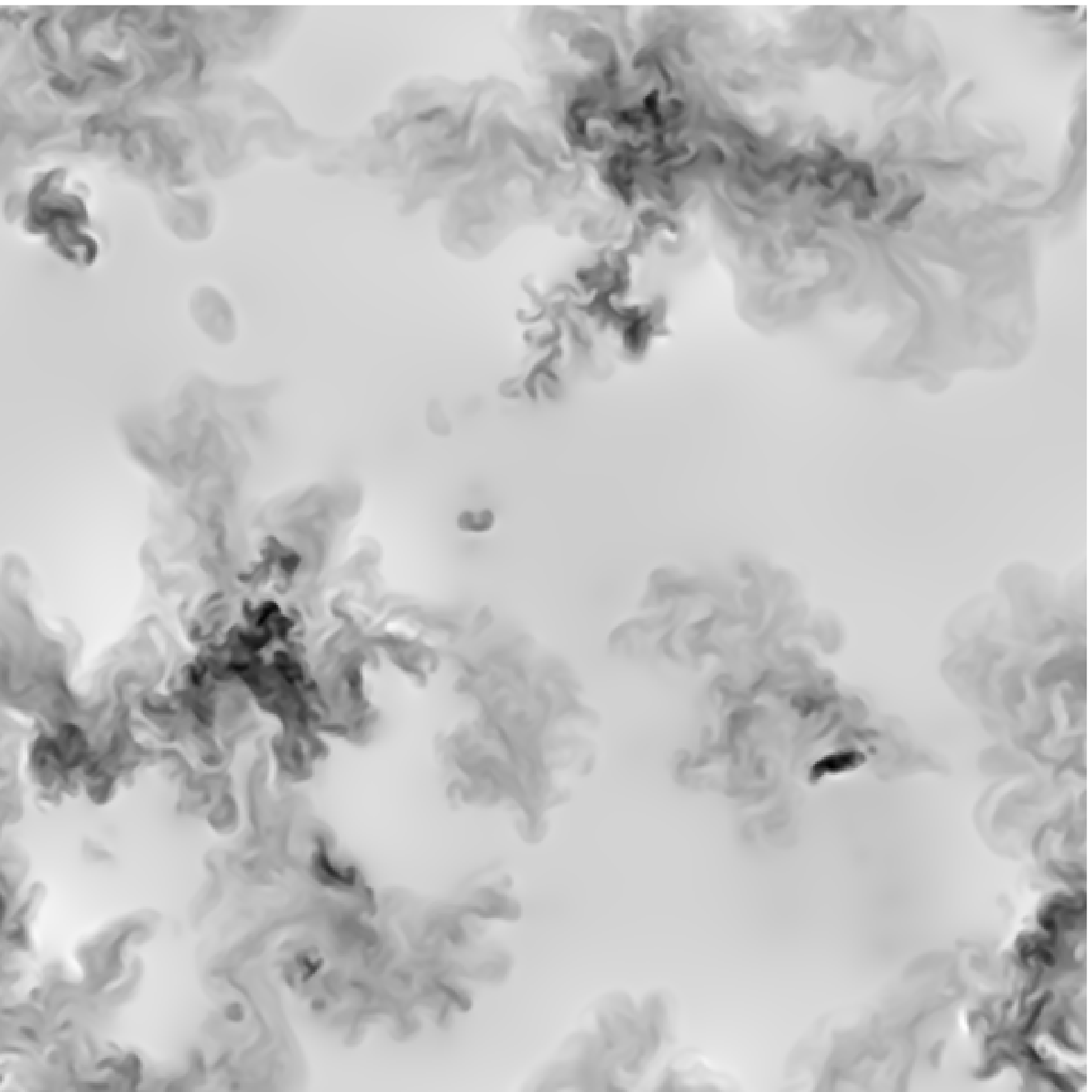}\hspace{2mm}
 \includegraphics[width=0.30\linewidth]{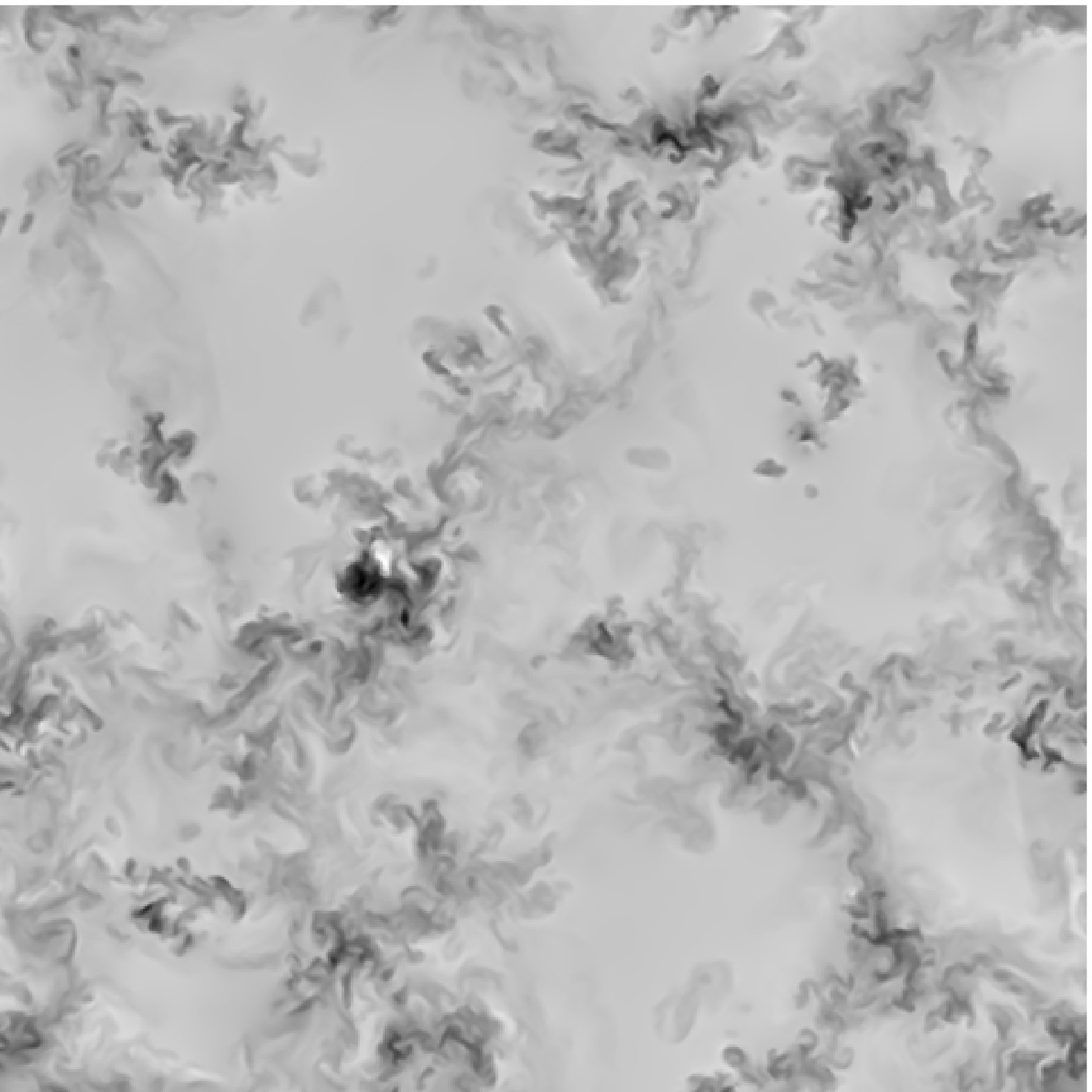}\hspace{2mm}
 \includegraphics[width=0.30\linewidth]{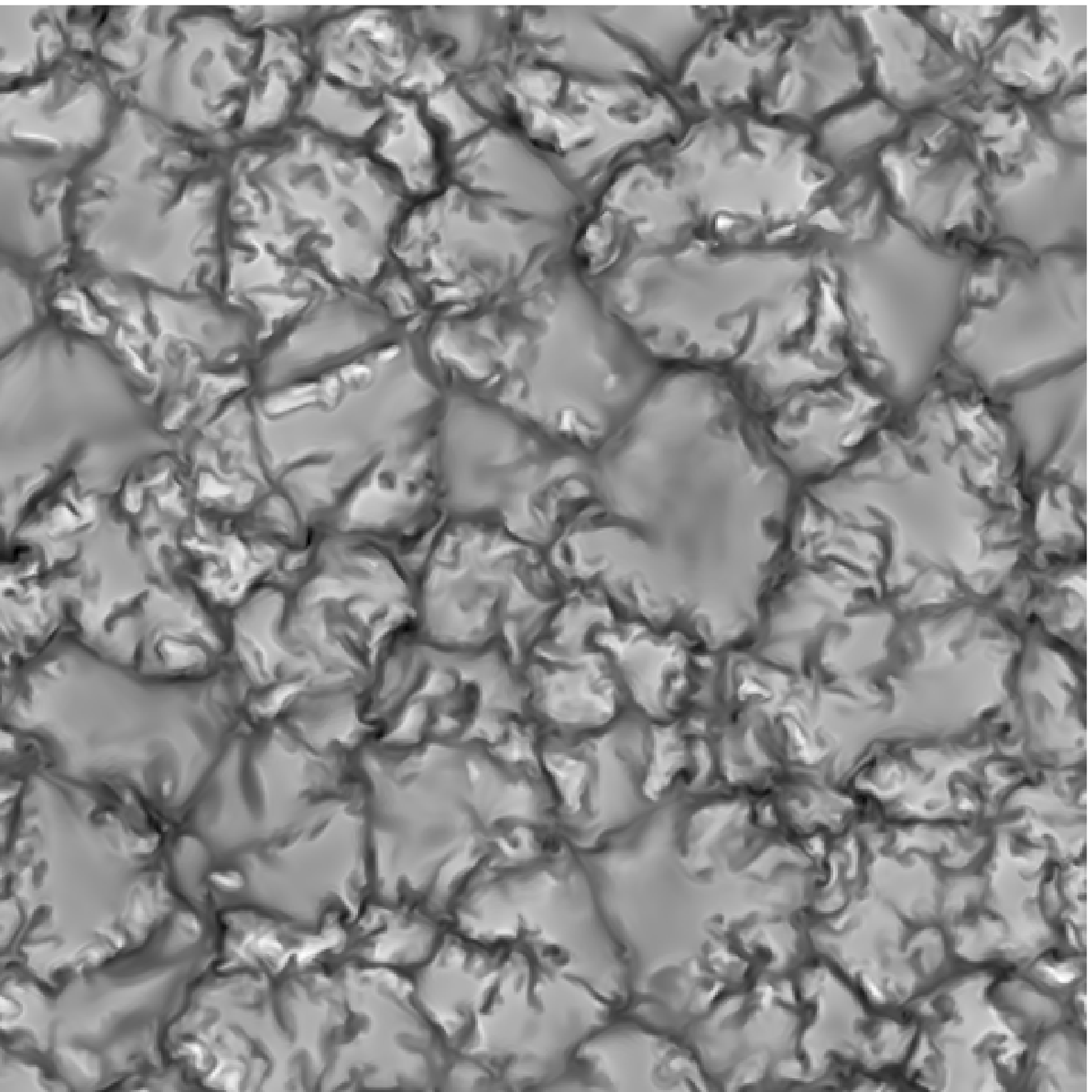}
 \caption{Snapshot from a simulation of small-scale dynamo action in the
          solar near-surface layers. The computational box is $12\times
          12\,$Mm$^2$ wide and $6.1\,$Mm deep. The panels show
          horizontal cuts of the vertical field component (upper panels;
          black and white indicates negative and positive polarity,
          respectively) and of the vertical flow velocity (lower panels;
          light shades indicates upflows, dark shades downflows). The
          cuts were taken at a depth of $\sim5$~Mm (left panels) and
          2.5~Mm (middle panels) below the average height of the optical
          surface, as well as at the optical surface (right panels). The
          dynamo-generated field is associated with the downflows in the
          deeper parts of the domain and thus exhibits a 'mesogranular'
          pattern at the surface.}
   \label{fig:dynamo}
\end{center}
\end{figure}

Fig.~\ref{fig:dynamo} shows a snapshot from a dynamo simulation carried
out in a deeper and wider computational box than previous
simulations. The maps show a close association between the
dynamo-generated field, which exhibits the characteristic mixed-polarity
pattern of small-scale dynamo action, and downflow areas. This is due to
the fact that the small-scale dynamo mainly works in the turbulent
downflows. The long-lived, large-scale horizontal convective flow
pattern in the deep layers is reflected in the 'mesogranular'
distribution of the magnetic field in the surface layers
\citep[cf.][]{Yelles:etal:2011}. The scale of this pattern is determined
by the relation between the horiontal advection time and the timescale
of flux generation (growth rate of the SSD) in the intergranular
downflows.

\section{Concluding remarks and outlook}
\label{sec:outlook}

The combined efforts of high-resolution observations and comprehensive
numerical simulations have tremendously improved our understanding of
the magneto-convective processes in the near-surface layers and the
photosphere of the Sun. The identification of key physical processes
through the simulations and the detailed quantitative comparison between
observational data and numerical results was the key factor for this
progress. It should be noted that such a comparison needs to be done in
a proper way, which means that observational data (e.g., intensity
images or Stokes profiles) should be compared with the corresponding
quantities derived from the simulation results (i.e., synthetic Stokes
profiles) after taking into account instrumental effects (e.g.,
convolving with point spread functions, adding straylight and noise,
rebinning to detector pixel size).  Comparing secondary
observational products such as inversion results directly with physical
quantities from a simulation can lead to misleading conclusions.

Where do we go from here, which are the frontiers of the field?
In my opinion, there are three lines of research that will dominate the
field in the coming years. 
\begin{enumerate}
\item The local comprehensive simulations will make contact to the
  global simulations of the deep convection zone carried out with
  anelastic codes. This will require wider and deeper boxes and the
  inclusion of the effects of sphericity and rotation. By combining the
  results (or even the codes), a comprehensive understanding of solar
  convection and its surface manifestations will hopefully be
  achieved. This is the more important considering recent helioseismic
  results that suggest an inconsistency with global simulations
  \citep{Hanasoge:etal:2010, Hanasoge:etal:2012, Gizon:Birch:2012}.
\item An second important goal is to extend the simulation domain into
  the chromosphere (and corona) with a proper treatment of the radiation
  processes. In the chromosphere, the energy balance is dominated by
  radiative transfer in a few strong spectral lines, all of which are
  formed in NLTE. In addition, ionization and recombination can no
  longer be treated as local equilibrium processes. A consistent
  treatment of these effects within the framework of a 3D MHD simulation
  provides a considerable challenge, which requires a significant amount
  of development work \citep[e.g.,][]{Gudiksen:etal:2011,
  Carlsson:Leenaarts:2012}. This also involves the treatment of the Hall
  effect and ambipolar diffusion in non-ideal MHD
  \citep[e.g.,][]{Cheung:Cameron:2012, Martinez-Sykora:etal:2012,
  Khomenko:Collados:2012}.

\item Finally, the comprehensive simulations have reached a degree of
  maturity and validation by solar observations that affords their
  application to other cool stars. 3D hydrodynamic simulations of this
  kind have been carried out already for some time
  \citep[e.g.,][]{Nordlund:Dravins:1990, Collet:etal:2007,
  Ludwig:etal:2009, Freytag:etal:2011}. Recently, the first
  magneto-convection simulations in the surface layers of other cool
  stars than the Sun have become available \citep{Beeck:etal:2011,
  Wedemeyer:etal:2012b}. Such simulations will provide a crucial tool
  for testing and calibrating methods to infer stellar magnetic fields
  from observational spectral and spectro-polarimetric data.
\end{enumerate}

\bibliography{schuessler.bbl}

\begin{thebibliography}{116}
\expandafter\ifx\csname natexlab\endcsname\relax\def\natexlab#1{#1}\fi

\bibitem[{{Abbett}(2007)}]{Abbett:2007}
{Abbett}, W.~P. 2007, \textit{ApJ}, 665, 1469

\bibitem[{{Abbett} \& {Fisher}(2012)}]{Abbett:Fisher:2012}
{Abbett}, W.~P. \& {Fisher}, G.~H. 2012, \textit{Sol. Phys.}, 277, 3

\bibitem[{{Attie} {et~al.}(2009){Attie}, {Innes}, \& {Potts}}]{Attie:etal:2009}
{Attie}, R., {Innes}, D.~E., \& {Potts}, H.~E. 2009, \textit{A\&A}, 493, L13

\bibitem[{{Beeck} {et~al.}(2012){Beeck}, {Collet}, {Steffen}, {Asplund},
  {Cameron}, {Freytag}, {Hayek}, {Ludwig}, \&
  {Sch{\"u}ssler}}]{Beeck:etal:2012}
{Beeck}, B., {Collet}, R., {Steffen}, M., {Asplund}, M., {Cameron}, R.~H.,
  {Freytag}, B., {Hayek}, W., {Ludwig}, H.-G., \& {Sch{\"u}ssler}, M. 2012,
  \textit{A\&A}, 539, A121

\bibitem[{{Beeck} {et~al.}(2011){Beeck}, {Sch{\"u}ssler}, \&
  {Reiners}}]{Beeck:etal:2011} {Beeck}, B., {Sch{\"u}ssler}, M., \&
  {Reiners}, A. 2011, in ASP Conf.
  Ser. 448, \textit{16th Cambridge Workshop on Cool Stars,
  Stellar Systems, and the Sun}, ed. C.~{Johns-Krull}, M.~K.
  {Browning}, \& A.~A. {West}, 1071

\bibitem[{{Bercik} {et~al.}(1998){Bercik}, {Basu}, {Georgobiani},
  {Nordlund}, \& {Stein}}]{Bercik:etal:1998} {Bercik}, D.~J., {Basu},
  S., {Georgobiani}, D., {Nordlund}, A., \& {Stein}, R.~F. 1998, in ASP
  Conf. Ser. 154: \textit{Cool Stars, Stellar Systems, and the Sun 10},
  ed. R.~A. {Donahue} \& J.~A. {Bookbinder} (San Francisco: Astronomical
  Society of the Pacific), 568

\bibitem[{{Bercik} {et~al.}(2003){Bercik}, {Nordlund}, \&
  {Stein}}]{Bercik:etal:2003} {Bercik}, D.~J., {Nordlund}, A., \&
  {Stein}, R.~F. 2003, in ESA Special Publication, Vol. 517,
  \textit{GONG+ 2002. Local and Global Helioseismology: the Present and
  Future}, ed. H.~{Sawaya-Lacoste}, 201--206

\bibitem[{{Bharti} {et~al.}(2010){Bharti}, {Beeck}, \&
  {Sch{\"u}ssler}}]{Bharti:etal:2010}
{Bharti}, L., {Beeck}, B., \& {Sch{\"u}ssler}, M. 2010, \textit{A\&A}, 510, A12

\bibitem[{{Bonet} {et~al.}(2008){Bonet}, {M{\'a}rquez}, {S{\'a}nchez Almeida},
  {Cabello}, \& {Domingo}}]{Bonet:etal:2008}
{Bonet}, J.~A., {M{\'a}rquez}, I., {S{\'a}nchez Almeida}, J., {Cabello}, I., \&
  {Domingo}, V. 2008, \textit{ApJ} (Letters), 687, L131

\bibitem[{{Bonet} {et~al.}(2010){Bonet}, {M{\'a}rquez}, {S{\'a}nchez Almeida},
  {Palacios}, {Mart{\'{\i}}nez Pillet}, {Solanki}, {del Toro Iniesta},
  {Domingo}, {Berkefeld}, {Schmidt}, {Gandorfer}, {Barthol}, \&
  {Kn{\"o}lker}}]{Bonet:etal:2010}
{Bonet}, J.~A., {M{\'a}rquez}, I., {S{\'a}nchez Almeida}, J., {Palacios}, J.,
  {Mart{\'{\i}}nez Pillet}, V., {Solanki}, S.~K., {del Toro Iniesta}, J.~C.,
  {Domingo}, V., {Berkefeld}, T., {Schmidt}, W., {Gandorfer}, A., {Barthol},
  P., \& {Kn{\"o}lker}, M. 2010, \textit{ApJ} (Letters), 723, L139

\bibitem[{Borrero \& Ichimoto(2011)}]{Borrero:Ichimoto:2011} Borrero,
J.~M. \& Ichimoto, K. 2011, \textit{Living Rev. Solar Phys.}, 8, 4\\
http://www.livingreviews.org/lrsp-2011-4

\bibitem[{{Brandenburg} \& {Dobler}(2002)}]{Brandenburg:Dobler:2002}
{Brandenburg}, A. \& {Dobler}, W. 2002, \textit{Comp. Phys. Comm.}, 147,
  471

\bibitem[{{Brandenburg} {et~al.}(2012){Brandenburg}, {Sokoloff}, \&
  {Subramanian}}]{Brandenburg:etal:2012}
{Brandenburg}, A., {Sokoloff}, D., \& {Subramanian}, K. 2012, \textit{Space Sci. Rev.}, 169, 123

\bibitem[{{Brandt} {et~al.}(1988){Brandt}, {Scharmer}, {Ferguson}, {Shine}, \&
  {Tarbell}}]{Brandt:etal:1988}
{Brandt}, P.~N., {Scharmer}, G.~B., {Ferguson}, S., {Shine}, R.~A., \&
  {Tarbell}, T.~D. 1988, \textit{Nature}, 335, 238

\bibitem[{{Brun} {et~al.}(2004){Brun}, {Miesch}, \& {Toomre}}]{Brun:etal:2004}
{Brun}, A.~S., {Miesch}, M.~S., \& {Toomre}, J. 2004, \textit{ApJ}, 614, 1073

\bibitem[{{Bushby} {et~al.}(2012){Bushby}, {Favier}, {Proctor}, \&
  {Weiss}}]{Bushby:etal:2012}
{Bushby}, P.~J., {Favier}, B., {Proctor}, M.~R.~E., \& {Weiss}, N.~O. 2012,
  \textit{Geoph. Astrophys. Fluid Dyn.}, 106, 508

\bibitem[{{Cameron} {et~al.}(2007){Cameron}, {Sch{\"u}ssler}, {V{\"o}gler}, \&
  {Zakharov}}]{Cameron:etal:2007b}
{Cameron}, R., {Sch{\"u}ssler}, M., {V{\"o}gler}, A., \& {Zakharov}, V. 2007,
  \textit{A\&A}, 474, 261

\bibitem[{{Carlsson} \& {Leenaarts}(2012)}]{Carlsson:Leenaarts:2012}
{Carlsson}, M. \& {Leenaarts}, J. 2012, \textit{A\&A}, 539, A39

\bibitem[{{Carlsson} {et~al.}(2004){Carlsson}, {Stein}, {Nordlund}, \&
  {Scharmer}}]{Carlsson:etal:2004}
{Carlsson}, M., {Stein}, R.~F., {Nordlund}, {\AA}., \& {Scharmer}, G.~B. 2004,
  \textit{ApJ} (Letters), 610, L137

\bibitem[{{Cattaneo}(1999)}]{Cattaneo:1999}
{Cattaneo}, F. 1999, \textit{ApJ}, 515, L39

\bibitem[{{Centeno}(2012)}]{Centeno:2012}
{Centeno}, R. 2012, \textit{ApJ}, 759, 72

\bibitem[{Charbonneau(2010)}]{Charbonneau:2010}
Charbonneau, P. 2010, \textit{Living Rev. Solar Phys.}, 7, 3,
  http://www.livingreviews.org/lrsp-2010-3

\bibitem[{{Cheung} \& {Cameron}(2012)}]{Cheung:Cameron:2012}
{Cheung}, M.~C.~M. \& {Cameron}, R.~H. 2012, \textit{ApJ}, 750, 6

\bibitem[{{Cheung} {et~al.}(2010){Cheung}, {Rempel}, {Title}, \&
  {Sch{\"u}ssler}}]{Cheung:etal:2010}
{Cheung}, M.~C.~M., {Rempel}, M., {Title}, A.~M., \& {Sch{\"u}ssler}, M. 2010,
  \textit{ApJ}, 720, 233

\bibitem[{Cheung {et~al.}(2007)Cheung, {Sch{\" u}ssler}, \&
  Moreno-Insertis}]{Cheung:etal:2007b}
Cheung, M.~C.~M., {Sch{\" u}ssler}, M., \& Moreno-Insertis, F. 2007, \textit{A\&A}, in
  press

\bibitem[{{Cheung} {et~al.}(2008){Cheung}, {Sch{\"u}ssler}, {Tarbell}, \&
  {Title}}]{Cheung:etal:2008}
{Cheung}, M.~C.~M., {Sch{\"u}ssler}, M., {Tarbell}, T.~D., \& {Title}, A.~M.
  2008, \textit{ApJ}, 687, 1373

\bibitem[{{Collet} {et~al.}(2007){Collet}, {Asplund}, \&
  {Trampedach}}]{Collet:etal:2007}
{Collet}, R., {Asplund}, M., \& {Trampedach}, R. 2007, \textit{A\&A}, 469, 687

\bibitem[{{Danilovic} {et~al.}(2010{\natexlab{a}}){Danilovic}, {Beeck},
  {Pietarila}, {Sch{\"u}ssler}, {Solanki}, {Mart{\'{\i}}nez Pillet}, {Bonet},
  {del Toro Iniesta}, {Domingo}, {Barthol}, {Berkefeld}, {Gandorfer},
  {Kn{\"o}lker}, {Schmidt}, \& {Title}}]{Danilovic:etal:2010a}
{Danilovic}, S., {Beeck}, B., {Pietarila}, A., {Sch{\"u}ssler}, M., {Solanki},
  S.~K., {Mart{\'{\i}}nez Pillet}, V., {Bonet}, J.~A., {del Toro Iniesta},
  J.~C., {Domingo}, V., {Barthol}, P., {Berkefeld}, T., {Gandorfer}, A.,
  {Kn{\"o}lker}, M., {Schmidt}, W., \& {Title}, A.~M. 2010{\natexlab{a}}, \textit{ApJ} (Letters),
  723, L149

\bibitem[{{Danilovic} {et~al.}(2012){Danilovic}, {R{\"o}hrbein}, {Cameron}, \&
  {Sch{\"u}ssler}}]{Danilovic:etal:2012}
{Danilovic}, S., {R{\"o}hrbein}, D., {Cameron}, R., \& {Sch{\"u}ssler}, M.
  2012, \textit{A\&A}, submitted

\bibitem[{{Danilovic} {et~al.}(2010{\natexlab{b}}){Danilovic}, {Sch{\"u}ssler},
  \& {Solanki}}]{Danilovic:etal:2010b}
{Danilovic}, S., {Sch{\"u}ssler}, M., \& {Solanki}, S.~K. 2010{\natexlab{b}},
  \textit{A\&A}, 513, A1

\bibitem[{{Deinzer} {et~al.}(1984){Deinzer}, {Hensler}, {Sch{\"u}ssler}, \&
  {Weisshaar}}]{Deinzer:etal:1984b}
{Deinzer}, W., {Hensler}, G., {Sch{\"u}ssler}, M., \& {Weisshaar}, E. 1984,
  \textit{A\&A}, 139, 426

\bibitem[{Fan(2009)}]{Fan:2009}
Fan, Y. 2009, \textit{Living Rev. Solar Phys.}, 6, 4,
  http://www.livingreviews.org/lrsp-2009-4

\bibitem[{{Fang} {et~al.}(2010){Fang}, {Manchester}, {Abbett}, \& {van der
  Holst}}]{Fang:etal:2010}
{Fang}, F., {Manchester}, W., {Abbett}, W.~P., \& {van der Holst}, B. 2010,
  \textit{ApJ}, 714, 1649

\bibitem[{{Fang} {et~al.}(2012){Fang}, {Manchester}, {Abbett}, \& {van der
  Holst}}]{Fang:etal:2012}
{Fang}, F., {Manchester}, IV, W., {Abbett}, W.~P., \& {van der Holst}, B. 2012,
  \textit{ApJ}, 745, 37

\bibitem[{{Freytag} {et~al.}(2012){Freytag}, {Steffen}, {Ludwig},
  {Wedemeyer-B{\"o}hm}, {Schaffenberger}, \&
  {Steiner}}]{Freytag:etal:2011} {Freytag}, B., {Steffen}, M., {Ludwig},
  H.-G., {Wedemeyer-B{\"o}hm}, S., {Schaffenberger}, W., \& {Steiner},
  O. 2012, \textit{J. Comp. Phys.}, 231, 919

\bibitem[{{Gilman} \& {Miller}(1981)}]{Gilman:Miller:1981}
{Gilman}, P.~A. \& {Miller}, J. 1981, \textit{ApJ Suppl.}, 46, 211

\bibitem[{{Gizon} \& {Birch}(2012)}]{Gizon:Birch:2012}
{Gizon}, L. \& {Birch}, A.~C. 2012, \textit{PNAS}, 109, 11896

\bibitem[{{Glatzmaier}(1985)}]{Glatzmaier:1985}
{Glatzmaier}, G.~A. 1985, \textit{ApJ}, 291, 300

\bibitem[{{Gudiksen} {et~al.}(2011){Gudiksen}, {Carlsson}, {Hansteen}, {Hayek},
  {Leenaarts}, \& {Mart{\'{\i}}nez-Sykora}}]{Gudiksen:etal:2011}
{Gudiksen}, B.~V., {Carlsson}, M., {Hansteen}, V.~H., {Hayek}, W., {Leenaarts},
  J., \& {Mart{\'{\i}}nez-Sykora}, J. 2011, \textit{A\&A}, 531, A154

\bibitem[{{Hanasoge} {et~al.}(2010){Hanasoge}, {Duvall}, \&
  {DeRosa}}]{Hanasoge:etal:2010}
{Hanasoge}, S.~M., {Duvall}, Jr., T.~L., \& {DeRosa}, M.~L. 2010, \textit{ApJ} (Letters), 712,
  L98

\bibitem[{{Hanasoge} {et~al.}(2012){Hanasoge}, {Duvall}, \&
  Sreenivasan}]{Hanasoge:etal:2012} {Hanasoge}, S.~M., {Duvall}, Jr.,
  T.~L., \& Sreenivasan, K.~R. 2012, \textit{PNAS}, 109, 11928

\bibitem[{{Heinemann} {et~al.}(2007){Heinemann}, {Nordlund}, {Scharmer}, \&
  {Spruit}}]{Heinemann:etal:2007}
{Heinemann}, T., {Nordlund}, {\AA}., {Scharmer}, G.~B., \& {Spruit}, H.~C.
  2007, \textit{ApJ}, 669, 1390

\bibitem[{{Jacoutot} {et~al.}(2008){Jacoutot}, {Kosovichev}, {Wray}, \&
  {Mansour}}]{Jacoutot:etal:2008}
{Jacoutot}, L., {Kosovichev}, A.~G., {Wray}, A., \& {Mansour}, N.~N. 2008,
  \textit{ApJ} (Letters), 684, L51

\bibitem[{{Joshi} {et~al.}(2011){Joshi}, {Pietarila}, {Hirzberger}, {Solanki},
  {Aznar Cuadrado}, \& {Merenda}}]{Joshi:etal:2011}
{Joshi}, J., {Pietarila}, A., {Hirzberger}, J., {Solanki}, S.~K., {Aznar
  Cuadrado}, R., \& {Merenda}, L. 2011, \textit{ApJ} (Letters), 734, L18

\bibitem[{{Keller} {et~al.}(2004){Keller}, {Sch{\" u}ssler}, {V{\" o}gler}, \&
  {Zakharov}}]{Keller:etal:2004}
{Keller}, C.~U., {Sch{\" u}ssler}, M., {V{\" o}gler}, A., \& {Zakharov}, V.
  2004, \textit{ApJ}, 607, L59

\bibitem[{{Khomenko} \& {Collados}(2012)}]{Khomenko:Collados:2012}
{Khomenko}, E. \& {Collados}, M. 2012, \textit{ApJ}, 747, 87

\bibitem[{{Kitiashvili} {et~al.}(2012{\natexlab{a}}){Kitiashvili},
  {Kosovichev}, {Mansour}, {Lele}, \& {Wray}}]{Kitiashvili:etal:2012b}
{Kitiashvili}, I.~N., {Kosovichev}, A.~G., {Mansour}, N.~N., {Lele}, S.~K., \&
  {Wray}, A.~A. 2012{\natexlab{a}}, \textit{Phys. Scr.}, 86, 018403

\bibitem[{{Kitiashvili} {et~al.}(2012{\natexlab{b}}){Kitiashvili},
  {Kosovichev}, {Mansour}, \& {Wray}}]{Kitiashvili:etal:2012a}
{Kitiashvili}, I.~N., {Kosovichev}, A.~G., {Mansour}, N.~N., \& {Wray}, A.~A.
  2012{\natexlab{b}}, \textit{ApJ} (Letters), 751, L21

\bibitem[{{Kitiashvili} {et~al.}(2010){Kitiashvili}, {Kosovichev}, {Wray}, \&
  {Mansour}}]{Kitiashvili:etal:2010}
{Kitiashvili}, I.~N., {Kosovichev}, A.~G., {Wray}, A.~A., \& {Mansour}, N.~N.
  2010, \textit{ApJ}, 719, 307

\bibitem[{{Lagg} {et~al.}(2010){Lagg}, {Solanki}, {Riethm{\"u}ller},
  {Mart{\'{\i}}nez Pillet}, {Sch{\"u}ssler}, {Hirzberger}, {Feller}, {Borrero},
  {Schmidt}, {del Toro Iniesta}, {Bonet}, {Barthol}, {Berkefeld}, {Domingo},
  {Gandorfer}, {Kn{\"o}lker}, \& {Title}}]{Lagg:etal:2010}
{Lagg}, A., {Solanki}, S.~K., {Riethm{\"u}ller}, T.~L., {Mart{\'{\i}}nez
  Pillet}, V., {Sch{\"u}ssler}, M., {Hirzberger}, J., {Feller}, A., {Borrero},
  J.~M., {Schmidt}, W., {del Toro Iniesta}, J.~C., {Bonet}, J.~A., {Barthol},
  P., {Berkefeld}, T., {Domingo}, V., {Gandorfer}, A., {Kn{\"o}lker}, M., \&
  {Title}, A.~M. 2010, \textit{ApJ} (Letters), 723, L164

\bibitem[{{Ludwig} {et~al.}(2009){Ludwig}, {Caffau}, {Steffen}, {Freytag},
  {Bonifacio}, \& {Ku{\v c}inskas}}]{Ludwig:etal:2009}
{Ludwig}, H.-G., {Caffau}, E., {Steffen}, M., {Freytag}, B., {Bonifacio}, P.,
  \& {Ku{\v c}inskas}, A. 2009, \textit{Mem. Soc. Astr. It.}, 80, 711

\bibitem[{{Mart{\'{\i}}nez-Sykora} {et~al.}(2012){Mart{\'{\i}}nez-Sykora}, {De
  Pontieu}, \& {Hansteen}}]{Martinez-Sykora:etal:2012}
{Mart{\'{\i}}nez-Sykora}, J., {De Pontieu}, B., \& {Hansteen}, V. 2012, \textit{ApJ},
  753, 161

\bibitem[{{Mart{\'{\i}}nez-Sykora} {et~al.}(2008){Mart{\'{\i}}nez-Sykora},
  {Hansteen}, \& {Carlsson}}]{Martinez:etal:2008}
{Mart{\'{\i}}nez-Sykora}, J., {Hansteen}, V., \& {Carlsson}, M. 2008, \textit{ApJ}, 679,
  871

\bibitem[{{Mart{\'{\i}}nez-Sykora} {et~al.}(2009){Mart{\'{\i}}nez-Sykora},
  {Hansteen}, \& {Carlsson}}]{Martinez:etal:2009}
---. 2009, \textit{ApJ}, 702, 129

\bibitem[{{Meneguzzi} {et~al.}(1981){Meneguzzi}, {Frisch}, \&
  {Pouquet}}]{Meneguzzi:etal:1981} {Meneguzzi}, M., {Frisch}, U., \&
  {Pouquet}, A. 1981, \textit{Phys. Rev. Let.}, 47, 1060

\bibitem[{{Moll} {et~al.}(2011{\natexlab{a}}){Moll}, {Cameron}, \&
  {Sch{\"u}ssler}}]{Moll:etal:2011b}
{Moll}, R., {Cameron}, R.~H., \& {Sch{\"u}ssler}, M. 2011{\natexlab{a}}, \textit{A\&A},
  533, A126

\bibitem[{{Moll} {et~al.}(2012){Moll}, {Cameron}, \&
  {Sch{\"u}ssler}}]{Moll:etal:2012}
---. 2012, \textit{A\&A}, 541, A68

\bibitem[{{Moll} {et~al.}(2011{\natexlab{b}}){Moll}, {Pietarila Graham},
  {Pratt}, {Cameron}, {M{\"u}ller}, \& {Sch{\"u}ssler}}]{Moll:etal:2011a}
{Moll}, R., {Pietarila Graham}, J., {Pratt}, J., {Cameron}, R.~H.,
  {M{\"u}ller}, W.-C., \& {Sch{\"u}ssler}, M. 2011{\natexlab{b}}, \textit{ApJ}, 736, 36

\bibitem[{{Muthsam} {et~al.}(2010){Muthsam}, {Kupka}, {L{\"o}w-Baselli},
  {Obertscheider}, {Langer}, \& {Lenz}}]{Muthsam:etal:2010}
{Muthsam}, H.~J., {Kupka}, F., {L{\"o}w-Baselli}, B., {Obertscheider}, C.,
  {Langer}, M., \& {Lenz}, P. 2010, \textit{New Astron.}, 15, 460

\bibitem[{{Nordlund}(1983)}]{Nordlund:1983} {Nordlund}, A. 1983, in
\textit{Solar Photosphere: Structure, Convection and Magnetic Fields},
IAU Symp. 138, ed. J.~O. {Stenflo} (Dordrecht: Reidel), 79

\bibitem[{{Nordlund}(1985{\natexlab{a}})}]{Nordlund:1985a}
{Nordlund}, A. 1985{\natexlab{a}}, \textit{Sol. Phys.}, 100, 209

\bibitem[{{Nordlund}(1985{\natexlab{b}})}]{Nordlund:1985b} {Nordlund},
A. 1985{\natexlab{b}}, in \textit{Theoretical Problems High Resolution
Solar Physics}, ed. H.~U. {Schmidt}, 101

\bibitem[{{Nordlund} \& {Dravins}(1990)}]{Nordlund:Dravins:1990}
{Nordlund}, A. \& {Dravins}, D. 1990, \textit{A\&A}, 228, 155

\bibitem[{{Nordlund} {et~al.}(2009){Nordlund}, {Stein}, \&
  {Asplund}}]{Nordlund:etal:2009} {Nordlund}, {\AA}., {Stein}, R.~F., \&
  {Asplund}, M. 2009, \textit{Living Rev. Solar Phys.}, 6, 2,\\
  http://www.livingreviews.org/lrsp-2009-2

\bibitem[{{Parker}(1963)}]{Parker:1963}
{Parker}, E.~N. 1963, \textit{ApJ}, 138, 552

\bibitem[{{Parker}(1978)}]{Parker:1978}
---. 1978, \textit{ApJ}, 221, 368

\bibitem[{{Pietarila Graham} {et~al.}(2010){Pietarila Graham}, {Cameron}, \&
  {Sch{\"u}ssler}}]{Pietarila:etal:2010}
{Pietarila Graham}, J., {Cameron}, R., \& {Sch{\"u}ssler}, M. 2010, \textit{ApJ}, 714,
  1606

\bibitem[{{Pietarila Graham} {et~al.}(2009){Pietarila Graham}, {Danilovic}, \&
  {Sch{\"u}ssler}}]{Pietarila:etal:2009}
{Pietarila Graham}, J., {Danilovic}, S., \& {Sch{\"u}ssler}, M. 2009, \textit{ApJ}, 693,
  1728

\bibitem[{{Rempel}(2011{\natexlab{a}})}]{Rempel:2011a}
{Rempel}, M. 2011{\natexlab{a}}, \textit{ApJ}, 729, 5

\bibitem[{{Rempel}(2011{\natexlab{b}})}]{Rempel:2011b}
---. 2011{\natexlab{b}}, \textit{ApJ}, 740, 15

\bibitem[{{Rempel}(2012)}]{Rempel:2012a}
---. 2012, \textit{ApJ}, 750, 62

\bibitem[{Rempel \& Schlichenmaier(2011)}]{Rempel:Schlichenmaier:2011}
Rempel, M. \& Schlichenmaier, R. 2011, \textit{Living Rev. Solar Phys.},
8, 3,\\ http://www.livingreviews.org/lrsp-2011-3

\bibitem[{{Rempel} {et~al.}(2009{\natexlab{a}}){Rempel}, {Sch{\"u}ssler},
  {Cameron}, \& {Kn{\"o}lker}}]{Rempel:etal:2009a}
{Rempel}, M., {Sch{\"u}ssler}, M., {Cameron}, R.~H., \& {Kn{\"o}lker}, M.
  2009{\natexlab{a}}, \textit{Science}, 325, 171

\bibitem[{{Rempel} {et~al.}(2009{\natexlab{b}}){Rempel}, {Sch{\"u}ssler}, \&
  {Kn{\"o}lker}}]{Rempel:etal:2009b}
{Rempel}, M., {Sch{\"u}ssler}, M., \& {Kn{\"o}lker}, M. 2009{\natexlab{b}},
  \textit{ApJ}, 691, 640

\bibitem[{{Robinson} {et~al.}(2003){Robinson}, {Demarque}, {Li}, {Sofia},
  {Kim}, {Chan}, \& {Guenther}}]{Robinson:etal:2003}
{Robinson}, F.~J., {Demarque}, P., {Li}, L.~H., {Sofia}, S., {Kim}, Y.-C.,
  {Chan}, K.~L., \& {Guenther}, D.~B. 2003, \textit{MNRAS}, 340, 923

\bibitem[{{R{\"o}hrbein} {et~al.}(2011){R{\"o}hrbein}, {Cameron}, \&
  {Sch{\"u}ssler}}]{Roehrbein:etal:2011}
{R{\"o}hrbein}, D., {Cameron}, R., \& {Sch{\"u}ssler}, M. 2011, \textit{A\&A}, 532, A140

\bibitem[{{S{\'a}nchez Almeida} \& {Mart{\'{\i}}nez
  Gonz{\'a}lez}(2011)}]{Sanchez:Martinez:2011}
{S{\'a}nchez Almeida}, J. \& {Mart{\'{\i}}nez Gonz{\'a}lez}, M. 2011, in
  ASP Conf. Ser. 437, \textit{Solar
  Polarization 6}, ed. J.~R. {Kuhn}, D.~M. {Harrington}, H.~{Lin}, S.~V.
  {Berdyugina}, J.~{Trujillo-Bueno}, S.~L. {Keil}, \& T.~{Rimmele}, 451

\bibitem[{{Schaffenberger} {et~al.}(2006){Schaffenberger},
  {Wedemeyer-B{\"o}hm}, {Steiner}, \&
  {Freytag}}]{Schaffenberger:etal:2006} {Schaffenberger}, W.,
  {Wedemeyer-B{\"o}hm}, S., {Steiner}, O., \& {Freytag}, B.  2006, in
  ASP Conf. Ser. 354, \textit{Solar MHD Theory and Observations: A High
  Spatial Resolution Perspective}, ed.  {J.~Leibacher, R.~F.~Stein, \&
  H.~Uitenbroek}, 345

\bibitem[{{Scharmer} {et~al.}(2011){Scharmer}, {Henriques}, {Kiselman}, \& {de
  la Cruz Rodr{\'{\i}}guez}}]{Scharmer:etal:2011}
{Scharmer}, G.~B., {Henriques}, V.~M.~J., {Kiselman}, D., \& {de la Cruz
  Rodr{\'{\i}}guez}, J. 2011, \textit{Science}, 333, 316

\bibitem[{{Sch{\"u}ssler} {et~al.}(2003){Sch{\"u}ssler}, {Shelyag},
  {Berdyugina}, {V{\"o}gler}, \& {Solanki}}]{Schuessler:etal:2003}
{Sch{\"u}ssler}, M., {Shelyag}, S., {Berdyugina}, S., {V{\"o}gler}, A., \&
  {Solanki}, S.~K. 2003, \textit{ApJ}, 597, L173

\bibitem[{{Sch{\"u}ssler} \& {V{\"o}gler}(2006)}]{Schuessler:Voegler:2006}
{Sch{\"u}ssler}, M. \& {V{\"o}gler}, A. 2006, \textit{ApJ} (Letters), 641, L73

\bibitem[{{Sch{\"u}ssler} \& {V{\"o}gler}(2008)}]{Schuessler:Voegler:2008}
---. 2008, \textit{A\&A}, 481, L5

\bibitem[{{Shelyag} {et~al.}(2011){Shelyag}, {Keys}, {Mathioudakis}, \&
  {Keenan}}]{Shelyag:etal:2011}
{Shelyag}, S., {Keys}, P., {Mathioudakis}, M., \& {Keenan}, F.~P. 2011, \textit{A\&A},
  526, A5

\bibitem[{{Shelyag} {et~al.}(2004){Shelyag}, {Sch{\" u}ssler}, {Solanki},
  {Berdyugina}, \& {V{\" o}gler}}]{Shelyag:etal:2004}
{Shelyag}, S., {Sch{\" u}ssler}, M., {Solanki}, S.~K., {Berdyugina}, S.~V., \&
  {V{\" o}gler}, A. 2004, \textit{A\&A}, 427, 335

\bibitem[{{Shelyag} {et~al.}(2007){Shelyag}, {Sch{\"u}ssler}, {Solanki}, \&
  {V{\"o}gler}}]{Shelyag:etal:2007}
{Shelyag}, S., {Sch{\"u}ssler}, M., {Solanki}, S.~K., \& {V{\"o}gler}, A. 2007,
  \textit{A\&A}, 469, 731

\bibitem[{{Spruit}(1976)}]{Spruit:1976}
{Spruit}, H.~C. 1976, \textit{Sol. Phys.}, 50, 269

\bibitem[{{Spruit}(1979)}]{Spruit:1979}
---. 1979, \textit{Sol. Phys.}, 61, 363

\bibitem[{{Spruit} \& {Zweibel}(1979)}]{Spruit:Zweibel:1979}
{Spruit}, H.~C. \& {Zweibel}, E.~G. 1979, \textit{Sol. Phys.}, 62, 15

\bibitem[{{Stein}(2012)}]{Stein:2012}
{Stein}, R.~F. 2012, \textit{Living Rev. Solar Phys.}, 9, 4,
  http://www.livingreviews.org/lrsp-2012-4

\bibitem[{{Stein} {et~al.}(2002){Stein}, {Bercik}, \&
  {Nordlund}}]{Stein:etal:2002}
{Stein}, R.~F., {Bercik}, D., \& {Nordlund}, A. 2002, Nuovo Cimento C
  \textit{Geophys. Space Phys. C}, 25, 513

\bibitem[{{Stein} {et~al.}(2011){Stein}, {Lagerfj{\"a}rd}, {Nordlund}, \&
  {Georgobiani}}]{Stein:etal:2011}
{Stein}, R.~F., {Lagerfj{\"a}rd}, A., {Nordlund}, {\AA}., \& {Georgobiani}, D.
  2011, \textit{Sol. Phys.}, 268, 271

\bibitem[{{Stein} \& {Nordlund}(1998)}]{Stein:Nordlund:1998}
{Stein}, R.~F. \& {Nordlund}, A. 1998, \textit{ApJ}, 499, 914

\bibitem[{{Stein} \& {Nordlund}(2006)}]{Stein:Nordlund:2006}
{Stein}, R.~F. \& {Nordlund}, {\AA}. 2006, \textit{ApJ}, 642, 1246

\bibitem[{{Stein} \& {Nordlund}(2012)}]{Stein:Nordlund:2012}
---. 2012, \textit{ApJ} (Letters), 753, L13

\bibitem[{{Steiner} {et~al.}(2010){Steiner}, {Franz}, {Bello Gonz{\'a}lez},
  {Nutto}, {Rezaei}, {Mart{\'{\i}}nez Pillet}, {Bonet Navarro}, {del Toro
  Iniesta}, {Domingo}, {Solanki}, {Kn{\"o}lker}, {Schmidt}, {Barthol}, \&
  {Gandorfer}}]{Steiner:etal:2010}
{Steiner}, O., {Franz}, M., {Bello Gonz{\'a}lez}, N., {Nutto}, C., {Rezaei},
  R., {Mart{\'{\i}}nez Pillet}, V., {Bonet Navarro}, J.~A., {del Toro Iniesta},
  J.~C., {Domingo}, V., {Solanki}, S.~K., {Kn{\"o}lker}, M., {Schmidt}, W.,
  {Barthol}, P., \& {Gandorfer}, A. 2010, \textit{ApJ} (Letters), 723, L180

\bibitem[{{Steiner} {et~al.}(1998){Steiner}, {Grossmann-Doerth}, {Kn\"olker},
  \& {Sch\"ussler}}]{Steiner:etal:1998}
{Steiner}, O., {Grossmann-Doerth}, U., {Kn\"olker}, M., \& {Sch\"ussler}, M.
  1998, \textit{ApJ}, 495, 468

\bibitem[{{Steiner} {et~al.}(1996){Steiner}, {Grossmann-Doerth}, {Schussler},
  \& {Knolker}}]{Steiner:etal:1996}
{Steiner}, O., {Grossmann-Doerth}, U., {Schussler}, M., \& {Knolker}, M. 1996,
  \textit{Sol. Phys.}, 164, 223

\bibitem[{{Steiner} \& {Rezaei}(2012)}]{Steiner:Rezaei:2012}
{Steiner}, O. \& {Rezaei}, R. 2012, ArXiv e-prints, arXiv:1202.4040v1

\bibitem[{{Trujillo Bueno} {et~al.}(2004){Trujillo Bueno}, {Shchukina}, \&
  {Asensio Ramos}}]{Trujillo:etal:2004}
{Trujillo Bueno}, J., {Shchukina}, N., \& {Asensio Ramos}, A. 2004, Nature,
  430, 326

\bibitem[{{Ustyugov}(2010)}]{Ustyugov:2010}
{Ustyugov}, S.~D. 2010, \textit{Physica Scripta Vol. T}, 142, 014031

\bibitem[{{V{\" o}gler}(2004)}]{Voegler:2004}
{V{\" o}gler}, A. 2004, \textit{A\&A}, 421, 755

\bibitem[{{V{\" o}gler} {et~al.}(2005){V{\" o}gler}, {Shelyag}, {Sch{\"
  u}ssler}, {Cattaneo}, {Emonet}, \& {Linde}}]{Voegler:etal:2005}
{V{\" o}gler}, A., {Shelyag}, S., {Sch{\" u}ssler}, M., {Cattaneo}, F.,
  {Emonet}, T., \& {Linde}, T. 2005, \textit{A\&A}, 429, 335

\bibitem[{{Vargas Dom{\'{\i}}nguez} {et~al.}(2011){Vargas Dom{\'{\i}}nguez},
  {Palacios}, {Balmaceda}, {Cabello}, \& {Domingo}}]{Vargas:etal:2011}
{Vargas Dom{\'{\i}}nguez}, S., {Palacios}, J., {Balmaceda}, L., {Cabello}, I.,
  \& {Domingo}, V. 2011, \textit{MNRAS}, 416, 148

\bibitem[{{V{\"o}gler}(2003)}]{Voegler:2003}
{V{\"o}gler}, A. 2003, PhD thesis, University of G{\"o}ttingen, Germany,\\
  http://webdoc.sub.gwdg.de/diss/2004/voegler (in English)

\bibitem[{{V\"ogler} \& {Sch\"ussler}(2003)}]{Voegler:Schuessler:2003}
{V\"ogler}, A. \& {Sch\"ussler}, M. 2003, Astron. Nachr./AN, 324, 399

\bibitem[{{V{\"o}gler} \& {Sch{\"u}ssler}(2007)}]{Voegler:Schuessler:2007}
{V{\"o}gler}, A. \& {Sch{\"u}ssler}, M. 2007, \textit{A\&A}, 465, L43

\bibitem[{{Webb} \& {Roberts}(1978)}]{Webb:Roberts:1978}
{Webb}, A.~R. \& {Roberts}, B. 1978, \textit{Sol. Phys.}, 59, 249

\bibitem[{{Wedemeyer} {et~al.}(2004){Wedemeyer}, {Freytag}, {Steffen},
  {Ludwig}, \& {Holweger}}]{Wedemeyer:etal:2004}
{Wedemeyer}, S., {Freytag}, B., {Steffen}, M., {Ludwig}, H.-G., \& {Holweger},
  H. 2004, \textit{A\&A}, 414, 1121

\bibitem[{{Wedemeyer} {et~al.}(2012){Wedemeyer}, {Ludwig}, \&
  {Steiner}}]{Wedemeyer:etal:2012b}
{Wedemeyer}, S., {Ludwig}, H.-G., \& {Steiner}, O. 2012, ArXiv e-prints,
  arXiv:1207.2342

\bibitem[{{Wedemeyer-B{\"o}hm}(2010)}]{Wedemeyer:2010}
{Wedemeyer-B{\"o}hm}, S. 2010, \textit{Mem. Soc. Astr. It.}, 81, 693

\bibitem[{{Wedemeyer-B{\"o}hm} {et~al.}(2012){Wedemeyer-B{\"o}hm}, {Scullion},
  {Steiner}, {Rouppe van der Voort}, {de La Cruz Rodriguez}, {Fedun}, \&
  {Erd{\'e}lyi}}]{Wedemeyer:etal:2012a}
{Wedemeyer-B{\"o}hm}, S., {Scullion}, E., {Steiner}, O., {Rouppe van der
  Voort}, L., {de La Cruz Rodriguez}, J., {Fedun}, V., \& {Erd{\'e}lyi}, R.
  2012, \textit{Nature}, 486, 505

\bibitem[{{Weiss}(1964)}]{Weiss:1964}
{Weiss}, N.~O. 1964, \textit{MNRAS}, 128, 225

\bibitem[{{Weiss}(1966)}]{Weiss:1966}
---. 1966, Proc. Roy. Soc. London, Ser. A, 293, 310

\bibitem[{{Weiss}(2012)}]{Weiss:2012}
---. 2012, \textit{Geoph. Astrophys. Fluid Dyn.}, 106, 353

\bibitem[{{Yelles Chaouche} {et~al.}(2009){Yelles Chaouche}, {Cheung},
  {Solanki}, {Sch{\"u}ssler}, \& {Lagg}}]{Yelles:etal:2009}
{Yelles Chaouche}, L., {Cheung}, M.~C.~M., {Solanki}, S.~K., {Sch{\"u}ssler},
  M., \& {Lagg}, A. 2009, \textit{A\&A}, 507, L53

\bibitem[{{Yelles Chaouche} {et~al.}(2011){Yelles Chaouche}, {Moreno-Insertis},
  {Mart{\'{\i}}nez Pillet}, {Wiegelmann}, {Bonet}, {Kn{\"o}lker}, {Bellot
  Rubio}, {del Toro Iniesta}, {Barthol}, {Gandorfer}, {Schmidt}, \&
  {Solanki}}]{Yelles:etal:2011}
{Yelles Chaouche}, L., {Moreno-Insertis}, F., {Mart{\'{\i}}nez Pillet}, V.,
  {Wiegelmann}, T., {Bonet}, J.~A., {Kn{\"o}lker}, M., {Bellot Rubio}, L.~R.,
  {del Toro Iniesta}, J.~C., {Barthol}, P., {Gandorfer}, A., {Schmidt}, W., \&
  {Solanki}, S.~K. 2011, \textit{ApJ} (Letters), 727, L30


\end{thebibliography}

\end{document}